\documentclass[fleqn,usenatbib]{mnras}
\usepackage{newtxtext,newtxmath}
\usepackage[T1]{fontenc}

\DeclareRobustCommand{\VAN}[3]{#2}
\let\VANthebibliography\thebibliography
\def\thebibliography{\DeclareRobustCommand{\VAN}[3]{##3}\VANthebibliography}

\usepackage{graphicx}
\usepackage{amsmath}

\newcommand{\AU}{\,{\rm au}}
\usepackage{color}
\definecolor{crevision}{rgb}{0.8, 0.2, 0.8}
\defcitealias{Martinez2021}{MB21}

\title[Signs of planet-disc interactions in V4046~Sgr]{The steady-state hydrodynamics of a long-lived disc: planetary system architecture and prospects of observing  a circumplanetary disc shadow in V4046~Sgr}

\author[P. Weber]{Philipp Weber$^{1,2,3}$
\thanks{Contact e-mail: \href{E-mail: philipppweber@gmail.com}{philipppweber@gmail.com}}
Simon Casassus,$^{1}$
and Sebasti{\'a}n P{\'e}rez,$^{2,3}$
\\
$^{1}$ Departamento de Astronom\'ia, Universidad de Chile, Camino El Observatorio 1515, Las Condes, Santiago, Chile\\
$^{2}$ Departamento de F\'isica, Universidad de Santiago de Chile, Av. Victor Jara 3659, Santiago\\
$^{3}$ Center for Interdisciplinary Research in Astrophysics and Space Exploration (CIRAS), Universidad de Santiago de Chile, Estaci\'on Central, Chile
}

\date{Accepted XXX. Received YYY; in original form ZZZ}

\pubyear{}

\begin{document}
\label{firstpage}
\pagerange{\pageref{firstpage}--\pageref{lastpage}}
\maketitle

\begin{abstract}
Recent imaging of the disc around the V4046~Sgr spectroscopic binary revealed concentric regions of dust rings and gaps.  The object's proximity and expected equilibrated state due to its old age (>20$\,$Myr) make it a superb testbed for hydrodynamical studies in direct comparison to observations. 
We employ two-dimensional hydrodynamical simulations of gas and multiple dust species to test whether the observed structure conforms with the presence of giant planets embedded in the disc. 
We then perform radiative transfer calculations of  sky images, which we filter for the telescope response for comparison with near-infrared and millimetre observations. 
We find that the existing data are in excellent agreement with a flared disc and the presence of two giant planets, at $9\AU$ and $20\AU$, respectively. The different ring widths are recovered by diffusion-balanced dust trapping within the gas pressure maxima. In our radiative transfer model, the diffusion in vertical direction is reduced in comparison to the radial value by a factor of five to recover the spectral energy distribution.
Further, we report a previously unaddressed, azimuthally-confined intensity decrement on the bright inner ring in the near-infrared scattered light observation. 
Our model shows that this decrement can be explained by a shadow cast by a circumplanetary disc around the same giant planet that creates the inner cavity in the hydrodynamical simulations. 
We examine the shape of the intensity indentation and discuss the potential characterisation of a giant planet and its associated disc by its projected shadow in scattered light observations.
\end{abstract}


\section{Introduction}
The technological improvement of telescopes in recent years has allowed the resolved observations of structures within protoplanetary discs (PPDs).
The optical and near-infrared (NIR) flux is dominated by stellar light scattered off small dust grains, and as PPDs are typically very optically thick to stellar radiation, it traces high altitudes in the disc's atmosphere. 
On the other hand, at longer wavelengths one typically observes colder and larger dust grains, expected to be located closer to the disc's mid-plane due to vertical settling. 
In this sense, observations by different instruments are complementary, as they trace grains of different sizes, and at different vertical heights in the disc. 
At high resolution, many intensity maps of PPDs show strong radial variations {(e.g. the DSHARP programme presented in \citealp{Andrews2018} and the DARTTS-S programme in \citealp{Avenhaus2018})}, presumably connected to rings and gaps in the dust density distributions. These structures are often associated with the presence of a planet in the depleted orbital zone {\citep[e.g.][]{Zhu2011,Rosotti2016,Zhang2018}}, yet this explanation is not unique. {Several alternative responsible mechanisms have been proposed: the creation of zonal flows due to variation of Maxwell stress in the presence of a vertical magnetic field \citep[][]{Johansen2009,Simon2014}, pile-up at the edge of dead-zones \citep[][]{Flock2015}, accumulation around ice lines \citep[][]{Saito2011,Banzatti2015,Zhang2015} or the combined effects of the snow line and presence of a magnetic field \citep[][]{Hu2019}.}
The detection of point-like signals from within {the depleted} regions can, however, strengthen the claim of a planetary origin. {Direct observation of protoplanets is challenging -- it is so far limited to the robust cases of PDS~70b \citep[detected at multiple infrared wavelengths,][]{Keppler2018,Muller2018} and PDS~70c \citep[detected from H$_\alpha$ emission,][]{Haffert2019}. Recently, the analysis of velocity perturbations in the gas kinematics due to the torque imposed by a giant planet has opened up an alternative indirect detection method based on the spectral shift of molecular emission lines \citep[][]{Perez2015,Perez2018,Pinte2018,Casassus2019,Pinte2020,Izquierdo2021}.}\\
\indent V4046 Sagittarii (also referred to as HD~319139 in early literature) is a double-lined spectroscopic binary system with an orbital period of 2.42~d \citep[corresponding to a semi-major axis of 0.041$\AU$,][]{Byrne1986,Quast2000}. It is at a distance of 71.48\,pc \citep{Gaia2020} and has an estimated combined binary mass of $m_{\rm b}\approx 1.75{\rm M}_\odot$ \citep[from the disc kinematics traced by $^{12}$CO$(2-1)$ line emission,][]{Rosenfeld2012} with a mass ratio close to unity \citep{Rosenfeld2013}. The binary stars are of K-type and are expected to be a part of the $\beta$ Pictoris moving group, whose median age was estimated to be $23\pm3\,$Myr \citep{Mamajek2014}. Assuming a value close to this median denotes an uncharacteristically old age for such substantial circumstellar material \citep[$\approx0.1 {\rm M}_\odot$,][]{Rosenfeld2013}. \\
\indent The system was imaged in the near-infrared (NIR) scattered light regime in several studies. Using J-band ($1.24\,\mu$m) and K2-band ($2.27\,\mu$m) filters of Gemini/GPI, \citet{Rapson2015} found a central cavity within $\approx 10\AU$, a relatively narrow ring of polarised NIR flux, centred at $\approx 15\AU$ and an adjacent gap at $\approx20\AU$. These structures were confirmed by polarimetric differential imaging (PDI) VLT/SPHERE observations in the J- and H-bands ($1.24\,\mu$m and $1.63\,\mu$m, respectively) with much higher signal-to-noise \citep{Avenhaus2018}.\\
\indent An accompanying article \citep[][subm., hereafter MB21]{Martinez2021} presents ALMA band 6 ($\lambda_{\rm obs}=1.25\,{\rm mm}$) observational data, obtained in Cycle 5 as part of the DARTTS-A programme. 
Using radiative transfer, the authors adapted a complex system of concentric rings, consisting of small and large dust grains, to VLT/SPHERE--IRDIS and ALMA band 6 data, as well as to the spectral energy distribution (SED). 
Compared to the NIR observations, the ALMA image shows a considerably thinner (yet resolved) ring of emission at $\approx13\AU$, consistent with the presence of a dust trap.\\
\indent {Explaining dust structures observed in high resolution at different wavelengths is a challenging endeavour. Attempts have been made, e.g. by \citet[][]{Ruiz-Rodriguez2019} for SPHERE observations and previous ALMA data of V4046~Sgr, by \citet[][]{Ballabio2021} for the system HD~143006 and \citet[][]{Brown-Sevilla2021} for WaOph~6. In the work at hand, we additionally take the SED into account, giving further stringent constraints for the system characteristics.} Given its age and apparent axial symmetry, V4046~Sgr is an ideal testbed for hydrodynamical models of planet-disc interaction in steady state, without the need to fine-tune transient features. Our intention in the work at hand is to propose a dynamical model able to explain the structures inferred in V4046~Sgr. We test the idea that the structures of the parametric model presented in \citetalias{Martinez2021} is due to the presence of planet-disc interaction. We reproduce the radial structure of both the scattered light and millimetre continuum images with radiative transfer predictions based on a hydrodynamical model including the presence of two giant planets. This adds to a previous study of the same system by \citet{Ruiz-Rodriguez2019} who compared observational data to the structures produced by a single giant planet embedded at $\approx 20\AU$.\\
\indent Next to the radial structure, the scattered light observations revealed interesting azimuthally-confined intensity decrements on the inner ring, linked to a local abatement of incident optical/infrared light. By a detailed analysis of both PDI and integral-field spectroscopic (IFS) data, \citet{DOrazi2019} were able to expose two such decrements as an effect of the inner binary's self-shadowing. Tracking those features across observational epochs allowed the authors to confirm the compliance of the movement of these shadows with the binary's period, as inferred from spectral lines \citep[][]{Quast2000}. In this work, we suggest the presence of an additional azimuthally-confined intensity decrement in the NIR. We discuss the possibility of observing the shadow of a circumplanetary disc (CPD) given the constraints of the existing data. \citet{Bertrang2020} and \citet{Montesinos2021} recently discussed a similar scenario for HD~169142, yet, in V4046~Sgr the reported feature appears much more prominent when compared against the azimuthal intensity variability along the projecting ring.

\section{Hydrodynamical Simulations}\label{sec:HD}
\subsection{Disc structure and evolution}
We introduce the PPD as a two-dimensional, vertically integrated model, depending on the radial coordinate, $r$, and the azimuth, $\varphi$. 
The circumbinary material is treated as locally isothermal, and consisting of one gas and multiple dust fluids where the number of $N_{\rm dust}$ different dust species account for the size dependency of the dust structure. The different dust species are marked by their corresponding index, $i=[1,N_{\rm dust}]$. For each fluid, the dynamically evolved quantities are the surface density, $\Sigma_{\rm g}$ for gas and $\Sigma_{\rm i}$ for dust, and the velocity vector, $\boldsymbol{u}$ for gas and $\boldsymbol{v}_{\rm i}$ for dust. The surface densities and velocities are evolved according to the continuity equations and momentum equations. The continuity equations for gas and dust read:
\begin{eqnarray}
\frac{\partial\Sigma_\mathrm{g}}{\partial t} + \boldsymbol{\nabla}\cdot \left(\Sigma_\mathrm{g}\boldsymbol{u} \right) \quad\,\,\,\,  =  0 \,, \label{eq:contgas} \\
\frac{\partial\Sigma_\mathrm{i}}{\partial t} + \boldsymbol{\nabla}\cdot \left(\Sigma_\mathrm{i}\boldsymbol{v}_{\rm i} +\boldsymbol{j}_{\rm i}\right)  =  0 \,, \label{eq:contdust}
\end{eqnarray}
where $\boldsymbol{j}_{\rm i}$ is the diffusive flux. Equations~(\ref{eq:contgas}) and (\ref{eq:contdust}) show that mass transfer between species is neglected in this treatment.
The momentum equations for gas and dust are:
\begin{eqnarray}
\Sigma_\mathrm{g}\left(\frac{\partial \boldsymbol{u}}{\partial t} + \boldsymbol{u}\cdot \mathbf{\nabla}\boldsymbol{u}\right) \,\,\,
 &=&   - \boldsymbol{\nabla}P - \boldsymbol{\nabla}\cdot\tau \; - \Sigma_\mathrm{g}\boldsymbol{\nabla} \phi - \Sigma_{\rm d} \boldsymbol{f}_{\rm d} \,,
\label{eq:NS-gas} \\[4pt]
\Sigma_\mathrm{d}\left(\frac{\partial \boldsymbol{v}_{\rm i}}{\partial t} + \boldsymbol{v}_{\rm i}\cdot \mathbf{\nabla}\boldsymbol{v}_{\rm i}\right)  &=&  -\Sigma_\mathrm{d}\boldsymbol{\nabla} \phi + \Sigma_{\rm d} \boldsymbol{f}_{\rm d}\,.
\label{eq:NS-dust}
\end{eqnarray}
In this work we use a locally isothermal gas pressure, $P=c_{\rm s}^2\Sigma_{\rm g}$, with a gas pressure scale height of $H_{\rm g} = c_{\rm s}\,\Omega_{\rm K}^{-1}$, where $c_{\rm s}$ denotes the speed of sound and $\Omega_{\rm K}$ the Keplerian frequency. The dust is assumed to be pressureless. Furthermore, the viscous stress tensor is given by $\tau \equiv \Sigma_{\rm g} \nu \left[ \boldsymbol{\nabla} \boldsymbol{u} + (\boldsymbol{\nabla}\boldsymbol{u})^T - \frac{2}{3}(\boldsymbol{\nabla}\cdot \boldsymbol{u})\boldsymbol{I}\right]$, with the identity matrix, $\boldsymbol{I}$. To describe the kinematic viscosity, $\nu$, we employ the $\alpha$-model \citep{Shakura1973} with a moderate level of viscosity, $\alpha=10^{-3}$.
Due to its small orbital separation, we approximate the central binary as a gravitational point source of mass $m_{\rm b}$ at the centre of the coordinate system, invoking the gravitational potential, $\phi_{\rm b} = -Gm_{\rm b}\,r^{-1}$. We neglect the disc's self-gravity. With $N_\mathrm{P}$ being an arbitrary number of planets in the system, the total gravitational potential is represented by $\phi = \phi_{\rm b} + \sum_{j}^{N_\mathrm{P}} \phi_{\mathrm{P},j}$. As our chosen coordinates are astrocentric (i.e. the radial coordinate is defined as the distance towards the central star and not towards the centre of gravity), a single planet's potential reads
\begin{equation}\label{eq:theo-planetpot}
\phi_\mathrm{P}(\boldsymbol{r}) = - \frac{Gm_{\mathrm{P}}}{|\boldsymbol{r}-\boldsymbol{r_{\mathrm{P}}}|} + \frac{Gm_\mathrm{P}}{r_\mathrm{P}^2} r \cos{\varphi}\,,
\end{equation}
where the second term arises exactly due to the acceleration of the aforementioned astrocentric frame.
The last term in equations~(\ref{eq:NS-gas}) and (\ref{eq:NS-dust}) accounts for the frictional interaction between gas and dust. In the Epstein regime, typically relevant for PPDs (and definitely for the work at hand), the frictional force per unit mass can be expressed by
\begin{equation}\label{eq:drag2}
    \boldsymbol{f}_{\rm i} = \frac{\Omega_{\rm K}}{\rm St_{\rm i}}(\boldsymbol{u} - \boldsymbol{v}_{\rm i})\,,
\end{equation}
with the dimensionless Stokes number
\begin{equation}\label{eq:stokes}
    {\rm St_{\rm i}} = \frac{\pi}{2} \frac{a_{\rm i}\rho_{\rm int}}{\Sigma_{\rm g}}\,.
\end{equation}
This number is, therefore, characteristic for the dynamical behaviour of a particle of size, $a_{\rm i}$, and intrinsic density, $\rho_{\rm int}$, at a certain location in the disc.
\begin{figure*}
    \centering
    \includegraphics[width=\textwidth]{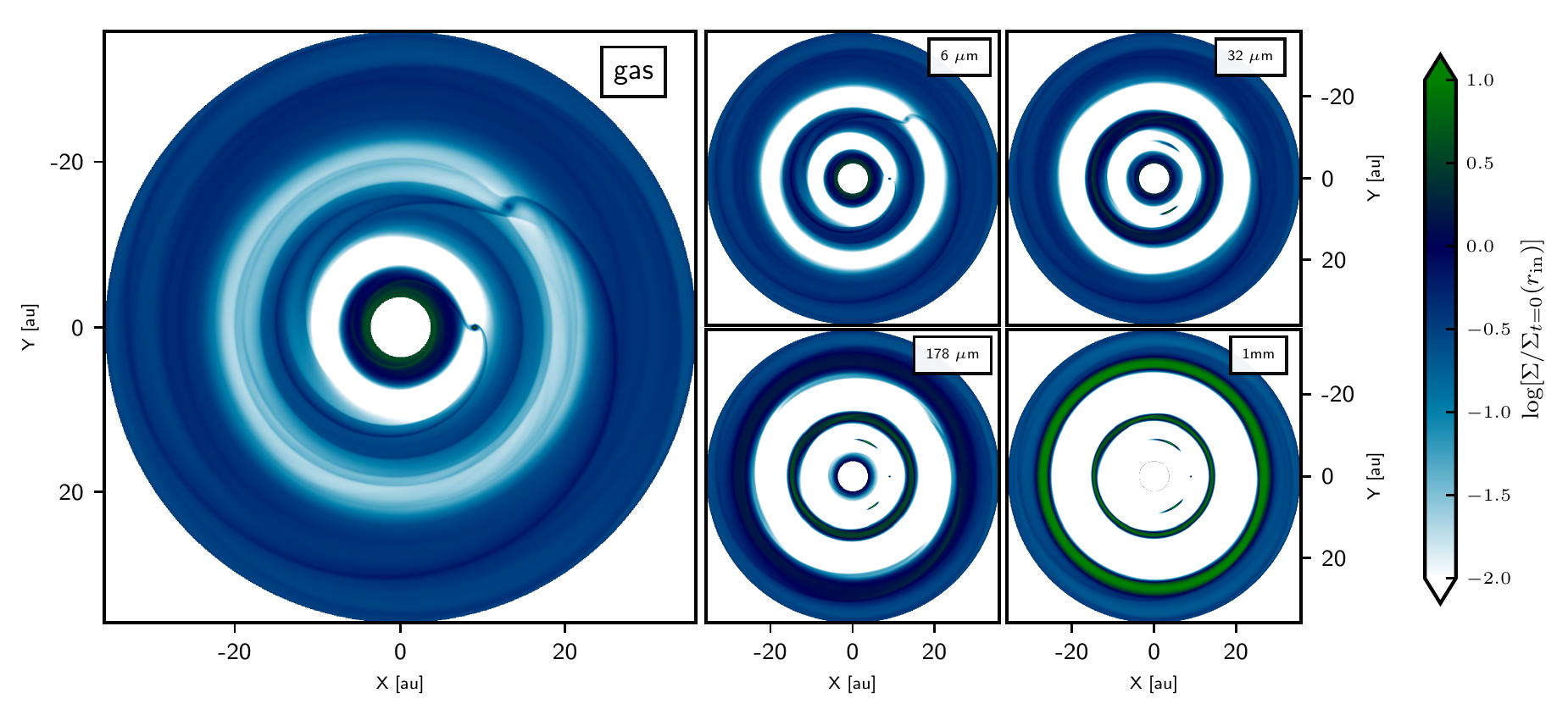}
    \vspace{-0.7cm}
    \caption{Logarithmic representation of gas and dust surface density distributions after 5000 orbits of P$_1$ at $9\AU$. To make them comparable, all the density fields are normalised by the corresponding initial value at the inner boundary of the grid. We show the surface density of the gas (left panel) and of four different dust species (right panels). The simulation includes two planets: P$_1$ of mass $m_{\rm P_2}=3.7\,{\rm M}_{\rm J}$ located at $r_{\rm P_1}=9\AU$ and P$_2$ of mass $m_{\rm P_2}=0.9\,{\rm M}_{\rm J}$ at $r_{\rm P2}=20\AU$.}
    \label{fig:Hydro results}
\end{figure*}

\subsection{Numerical Setup}
For our two-dimensional hydrodynamical simulations, we employ the publicly available code FARGO3D\footnote{\hyperlink{http://fargo.in2p3.fr/}{http://fargo.in2p3.fr/}} \citep{PBLL2016}, modelling dust as multiple pressureless fluids \citep{PBLL2019} and making use of the FARGO algorithm \citep{Masset2000}. We additionally allow the dust concentration to diffuse, as described in the appendix of \citet{Weber2019}.

We start the simulation from an unperturbed initial condition of a PPD, defined by radial power-law profiles for densities and temperature: 
\begin{equation}\label{equ:power-laws}
    \qquad \qquad \Sigma = \Sigma_0 \left(\frac{r}{r_0}\right)^{-1}\,, \qquad
    T=T_0\left(\frac{r}{r_0}\right)^{-0.5}\,.
\end{equation}
This implies a flared disc {(described by the flaring angle, $f=1.25$)}, where the aspect ratio, $h=H_{\rm g}/r= h_0(r/r_0)^{f-1}$, increases with the distance to the star. 
{The power-law exponents are chosen to initiate a disc in steady-state, approximately consistent with slopes in \citet[][]{Flaherty2020} ($\Sigma \propto r^{-1}, T\propto r^{-0.56}$), \citet[][]{DOrazi2019} ($f\sim 1.5$), \citet[][]{Avenhaus2018} ($f \sim 0.093-1.30$) and \citetalias{Martinez2021} ($\Sigma \propto r^{-1},\, f\sim 1.2 - 1.5$). The surface density exponent is consistent with theoretical values for an $\alpha$-model of a gaseous disc \citep[][]{Bell1997}.}
We set $h_0=0.035$, describing a relatively thin disc. 
For the gas density, we set the value $\Sigma_{\rm g,0}=10\,{\rm g}\,{\rm cm}^{-2}$ at $r_0=10\AU$. The dust densities are then calculated by demanding that locally the total dust-to-gas ratio is 0.01 and by employing a dust size distribution for the number density of $n(a)\propto a^{-3.5}$. In our simulation, we sample the size distribution by five dust fluids with sizes logarithmically-spaced from $a_{\rm min}=1\,\mu$m to $a_{\rm max}=1\,{\rm mm}$. This signifies that $\Sigma_{\rm i} \propto a^{0.5}$, under the consideration that on a logarithmically-spaced grid larger dust sizes need to account for a larger bin size. 

The parametric model presented in \citetalias{Martinez2021} predicts three dust rings, at roughly $6\AU$, $13\AU$ and $24\AU$, henceforth R$_1$, R$_2$ and R$_3$, respectively. To explain this structure with the presence of planets, we expect the system to host at least two giant planets, P$_1$ and P$_2$. In this light, P$_1$ is expected to create the separation between R$_1$ and R$_2$ and P$_2$ the gap between R$_2$ and R$_3$. We use planetary masses of $m_{{\rm P}_1}=3.7\,{\rm M}_{\rm J}$ and $m_{{\rm P}_2}=0.9\,{\rm M}_{\rm J}$, corresponding to fractions of the binary mass of $q_1=2\times10^{-3}$ and $q_2=5\times10^{-4}$. The locations of the planets are set to $r_{\rm P_1} = 9\AU$ and $r_{\rm P_2}=20\AU$.

We numerically evolve the system on a grid of $N_r \times N_\varphi = 512 \times 1024$ with logarithmic radial spacing between $4\AU<r<40\AU$, linearly spread over the full azimuthal domain. 
\begin{figure}
    \centering
    \includegraphics[width=\columnwidth]{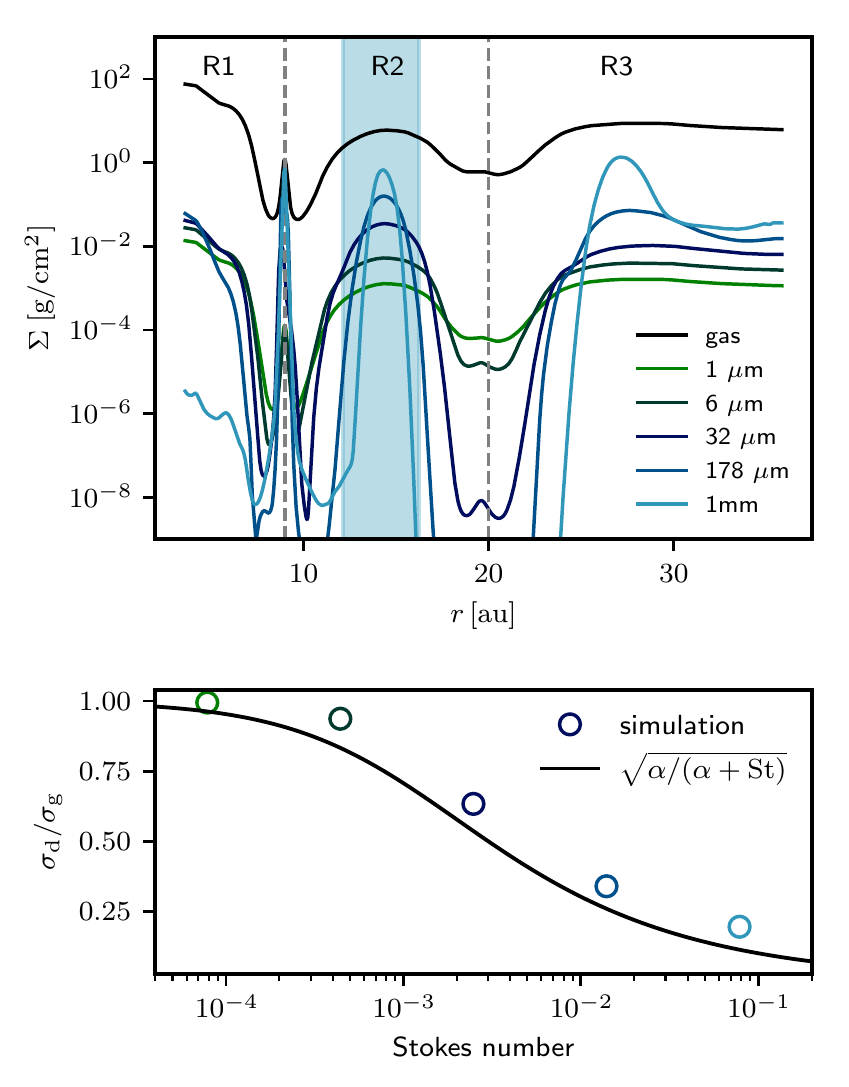}
    \vspace{-0.7cm}
    \caption{The \textit{top panel} shows the azimuthally-averaged density distribution including gas and five different dust species after 5000 orbits of P$_1$. The vertical dashed lines mark the positions of the two included planets at $r_{\rm P_1}=9\AU$ and $r_{\rm P_2}=20\AU$. The ring R$_2$ was fitted with Gaussians in the range of the shaded area. In the \textit{bottom panel}, the circles compare the obtained widths for different dust species (dependent on their Stokes number) to the width of the gas ring. The colours of the circles correspond to the legend of the top panel. The black solid line marks the expectation from theoretical considerations of dust trapping.}
    \label{fig:1D}
\end{figure}

\subsection{Results of hydrodynamical simulation}
\subsubsection{Density distribution}
The surface density fields of gas and dust are portrayed in Fig.~\ref{fig:Hydro results} after 5000 orbits of P$_1$ and normalised by the corresponding density value at the inner boundary of the initial condition (equation~\ref{equ:power-laws}).
Fig.~\ref{fig:Hydro results} shows that the two included planets (P$_1$ and P$_2$) form two significant gaps and enclose a radially-confined ring. The different panels on the right also show that for larger grains the gap areas are wider and the inner system becomes depleted. Qualitatively, the radial structure becomes better visible in Fig.~\ref{fig:1D}. The top panel shows the azimuthally-averaged surface density profiles. Subsequently, for each fluid we fitted a Gaussian to the ring distribution centred at $\approx 14\AU$, considering only data points within the shaded area (labeled R$_2$). In hydrostatical equilibrium the width of the dust ring formed by a certain size, $\sigma_{\rm d}(a_{\rm i})$, depends on the width of the sub-lying gas ring, $\sigma_{\rm g}$, via the connection, $\sigma_{\rm d}(a_{\rm i})=\sigma_{\rm g}\sqrt{D/(D+{\rm St}(a_{\rm i}))^{-1}}$ \citep[e.g.][]{Dubrulle1995,Dullemond2018}, where $D$ is a dimensionless parameter characterising the strength of diffusivity. In our simulations, it is set equal to the level of turbulent viscosity, $\alpha$. To estimate the Stokes number corresponding to a certain size, we use the gas surface density at the ring peak and insert it into equation~(\ref{eq:stokes}). In the bottom panel of Fig.~\ref{fig:1D} one can see that the general trend expected from theoretical studies of dust trapping can be recovered, yet the width of the dust rings is systematically slightly larger in the simulation.

Larger dust grains (32$\mu$m, 178$\mu$m, 1mm) get additionally trapped in the co-orbital region of planet P$_1$, at Lagrange points L$_4$ and $L_5$. We measure the respective trapped dust masses to be 0.09$\,$m$_{\oplus}$ and 0.18$\,$m$_{\oplus}$, respectively. This trapping has been studied in details by \citet[][]{Montesinos2020} with a focus on the origin of the Trojans in the Solar System and by \citet[][]{Rodenkirch2021} with a focus on the observability by ALMA as e.g. suggested in the system HD~163296. Both works show that the accumulated mass declines with time and that turbulent viscosity promotes the evacuation of the Lagrange points. The observed asymmetry of trapped dust masses (favouring L$_5$) in our study is consistent with both dedicated works.

{Recapitulating, the chosen hydrodynamical setup} qualitatively fulfils the characteristics of V4046~Sgr as reported in \citetalias{Martinez2021}: a {dust reservoir radially inwards of P$_2$} where large grains are depleted, a ring at $\approx14\AU$ that becomes more confined with increasing grain size and a gap at $\approx20\AU$ adjacent to the outer disc.

\section{Synthetic Observations}\label{sec:obs}
To calculate the observable fluxes arising from the hydrodynamical simulation we use the publicly available code RADMC-3D\footnote{\hyperlink{https://www.ita.uni-heidelberg.de/~dullemond/software/radmc-3d/}{https://www.ita.uni-heidelberg.de/~dullemond/software/radmc-3d/}} \citep[version2.0,][]{Dullemond2012}.
\subsection{Radiative Transfer Model}\label{subsec:RTmodel}
We implement the parameters of V4046~Sgr binary  into the radiative transfer (RT). We use the same values as in \citetalias{Martinez2021} for the stellar parameters, thus almost equally weighted components, $m_{\star,1} = 0.9\,{\rm M}_\odot$ and $m_{\star,2} = 0.85\,{\rm M}_\odot$, radii of $R_{\star,1}=1.064\,{\rm R}_\odot$ and $R_{\star,2}=1.033\,{\rm R}_\odot$ and temperatures of $T_{\star,1}=4350\,$K and $T_{\star,2}=4060\,$K.  
For the combined accretion rate we use a value of $\dot{m}_{\rm b} = 10^{-9.3} {\rm M}_\odot {\rm yr}^{-1}$ \citep{Donati2011}, and $a_{\rm b}=0.041\AU$ for their orbital separation. The stars are modelled as two Kurucz photospheres \citep{Kurucz1979} and implemented as extended light sources. The distance to the binary is taken to be $d=71.48\,$pc \citep{Gaia2020} and the system's inclination, $i=32.96^\circ$, and position angle, ${\rm PA}=77.31^\circ$ \citepalias{Martinez2021}.  

The density fields of the hydrodynamic simulations form the basis of the RT calculations. Additionally to the five dust species of the simulation's output, we expand the grain size distribution to $a_{\rm min}=0.2\,\mu$m in the RT to account for scattering at short wavelengths. To do so we use the density field of the smallest size that was included in the hydro ($a_{\rm min}=1\,\mu$m) and scale it to fit into the grain size distribution. This simple expansion of the size distribution is justified, as the dynamics of grains with sizes $<1\,\mu$m are expected to be similarly tightly coupled to the gas motion and their dynamical feedback onto the gas structure is negligible due to the low contribution to the total dust mass. The distribution is truncated for values smaller than $a_{\rm min}=0.2\,\mu$m to fit the observed NIR asymmetry as will be discussed in Section~\ref{subsubsec:scatterlight}.

We then expand the polar grid to three-dimensional, spherical coordinates, with an opening angle of $\Delta \theta=\pm\pi/4$ in both directions of the mid-plane, sampled by 65 cells that are logarithmically refined around the mid-plane. We spread the dust densities vertically by transforming each surface density, $\Sigma_{\rm i}$, into a three-dimensional volume density, $\rho_{\rm i}$, following
\begin{equation}\label{equ:vertical}
    \rho_{\rm i} = \frac{\Sigma_{\rm i}(r,\varphi)}{\sqrt{2\pi} H_{\rm i}} \exp{\left(-\frac{z^2}{2H_{\rm i}^2}\right)}\,,
\end{equation}
where $z\equiv r\cos{\theta}$ and the vertical dust scale height 
\begin{equation}
    H_{\rm i} = \sqrt{\frac{D_z}{D_z+{\rm St_{\rm i}}}}H_{\rm g}\,,
\end{equation}
with the vertical diffusion coefficient, $D_z$.
Note that this expression is equivalent to the relation between gas and dust ring widths (given in Section~\ref{sec:HD}) as the same dynamical processes prevail: the diluting effect of dust diffusion balanced by settling resulting from frictional coupling to a Gaussian background gas density.
In a similar manner as for the ring widths, it produces sharper confinement for larger grains, while smaller grains mirror the gas' vertical profile.
We find that in order to reproduce observed ALMA fluxes and the spectral energy distribution (SED, see Appendix~\ref{appendix:SED}), the vertical diffusion needs to be reduced compared to the $\alpha$-viscosity value. We use $D_z = 0.2\times\alpha$ in our RT model.
For the radial and azimuthal coordinates we use the same grid as in the hydrodynamical simulations but append 64 cells at the outer boundary extending to $65\AU$ to extrapolate the disc size using the initial power-law slopes for surface density and temperature.
To calculate the SED, we set up a wavelength grid between $0.1$~$\mu$m$<\,\lambda<2000$~$\mu$m, sampled by 200 logarithmically-spaced values. 
We do not analyse the SED further here but used it as a tool to increase the quality of our model at all wavelengths. The comparison to observational data is shown in Appendix~\ref{appendix:SED}.
We assume a dust composition of 60$\,$per cent silicates, 20$\,$per cent water ice and 20$\,$per cent amorphous carbon and compute the corresponding dust opacities with the {\tt bhmie} code provided in the RADMC-3D package. 
Finally, we use $10^8\,$photon packages to calculate the disc temperature, $10^8\,$photon packages to trace the thermal emission at a wavelength of 1.25~mm, and $2\times10^8\,$packages of scattering photons to obtain a synthetic image in the NIR. 
To account for the angle-dependent scattering relevant for the presented case, we use the RADMC-3D option {\tt scattering\_mode\_max=5} which includes the full anisotropic treatment (considering randomly distributed, spherical dust grains) by calculating the Müller scattering matrix according to \citet[][]{Bohren1983}\footnote{RADMC-3D includes a python transcript of the Draine-version (\hyperlink{http://scatterlib.wikidot.com/mie}{http://scatterlib.wikidot.com/mie}).}. This treatment reproduces the observed effect of forward-peaked scattering.

Finally, our hydrodynamical simulations show that the flux through the planetary gap is halted for dust sizes of $a\gtrsim 100\,\mu$m and those sizes are filtered out from the inner system. The present dust population of large grains in the inner cavity is expected to be evacuated when evolving the disc for longer. We hence deplete the inner cavity manually for these sizes.

Additionally, we implement an inner circumbinary disc between $0.17\AU < r < 0.8\AU$ to reproduce the centrally confined emission found in the ALMA observations and to improve the fit to the SED at intermediate wavelengths.

\subsection{Results and telescope responses}
In the following, we present the results of the RT calculations for different observational features and shortly discuss their significance.  

\subsubsection{Scattered light Images}\label{subsubsec:scatterlight}
\begin{figure}
    \centering
    \includegraphics[width=\columnwidth]{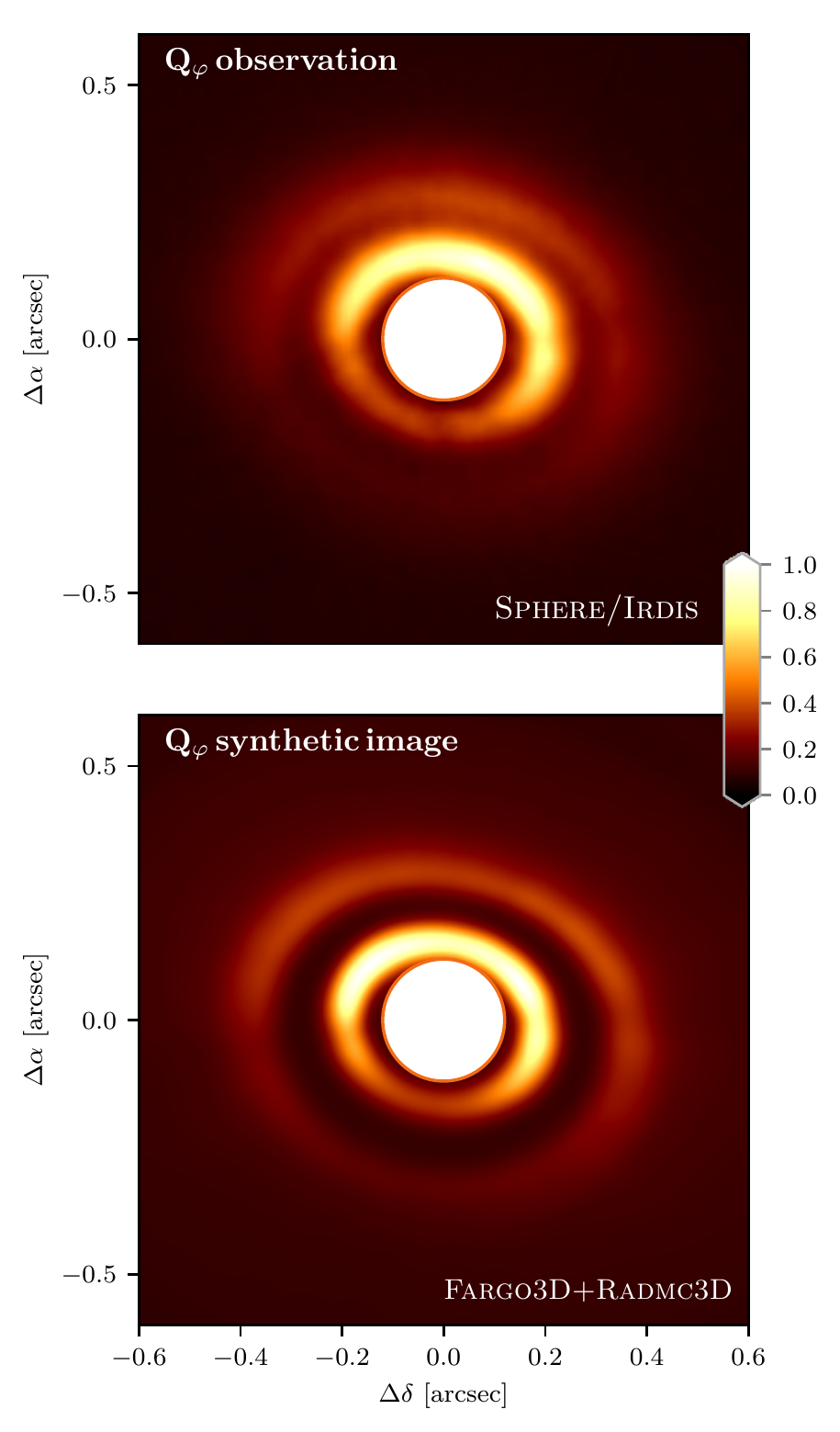}
    \vspace{-0.5cm}
    \caption{Comparison of observations and synthesis. {The colourbar shows the intensity on linear scale, normalised to the peak value.} The \textit{top panel} shows VLT/SHPERE H-band data taken by \citet{Avenhaus2018} (13$^{\rm th}$ of March 2016) after additional IRDAP reduction \citepalias[presented in ][]{Martinez2021}. The \textit{bottom panel} shows the results of our hydrodynamical model and RT. We smoothed the image with a point spread function (PSF, 0.05$\,$arcsec) tailored to reproduce the observational conditions of the telescope.}
    \label{fig:VLT}
\end{figure}
Radiation that is scattered towards the observer is expected to be linearly polarised perpendicularly to the incoming, pre-scattering direction. In the NIR the observations are typically dominated by stellar light scattered off small particles in the upper layers of the PPD. Under the assumption that each observed photon only participated in one single scattering event, the ray of incident radiation points towards the central binary \textit{independent of the location in the disc}, and the scattered light is expected to be mostly linearly polarised in the azimuthal direction, perpendicular to the direction of incident light. Therefore, we consider the polarised fluxes in their polar representation, with:
\begin{equation}
    Q_\varphi = +Q \cos(2\varphi) + U \sin(2\varphi)\,,
\end{equation}
where $Q$ and $U$ are components of the Stokes vector. This representation is equivalent to the choice in \citet{Avenhaus2018}.
The left panel of Fig.~\ref{fig:VLT} shows the exisiting PDI data of \citealp{Avenhaus2018}, re-reduced with the IRDAP pipeline \citep{vanHolstein2020} as presented in \citetalias{Martinez2021}, the right panel shows our corresponding synthetic image. 
The synthetic image of polarised scattered light in Fig.~\ref{fig:VLT} highlights further that even though the dust density distributions do not depend much on the azimuth, the observed scattered light shows a strong discrepancy between the northern and southern halves. As discussed in \citetalias{Martinez2021}, this is consistent with a strongly forward-peaked scattering function, resulting in a brighter near-side and a fainter far-side of the disc. This contrast is lost when lowering the minimum grain size to a regime in which Mie-scattering is not predominant and, therefore, points to an evolved dust population (to produce the contrast observed in the bottom panel of Fig.~\ref{fig:VLT} the minimum grain size was set to 0.2$\,\mu$m).

\subsubsection{ALMA continuum image}
The synthetic continuum image at $\lambda_{\rm obs} = 1.25\,$mm is compared to the observation \citepalias{Martinez2021} in Fig.~\ref{fig:ALMA}. For this we performed image synthesis (using CASA, cycle 5.8 configuration\footnote{The real observation was performed in C43-8/9 configuration, yet, the {\tt uvmem} image synthesis leads to a beam size better represented by the cycle 5.8 configuration.}) onto the RADMC-3D output, choosing a similar setup as for the real observations. To make the two images directly comparable, the intensity of the synthetic image is re-scaled to the \textit{observational} beam size. Apart from the unaccounted central emission, the employed model reproduces all the important features of the real observation. The ring R$_2$ between the two outer planetary orbits is very radially confined, also the intensity structure in the outer disc is well recovered. The accumulations of Large dust grains at the Lagrange points L$_4$ and L$_5$ are not observable with this setup.

\section{Discussion}
\subsection{Hydrodynamics and RT}
The analysis of the system V4046~Sgr is of specific interest as it constitutes one of the closest PPDs and offers variable observational data, such as NIR scattered light, ALMA continuum and the SED, that all show interesting structure.

As mentioned in Sec.~\ref{subsubsec:scatterlight}, the global intensity variation between the near- and far-side seen in scattered light is linked to forward-peaked scattering. In our model we assumed dust grains to be compact and spherical at all sizes.
As discussed in \citet[][]{Stolker2016}, for such grains the global intensity contrast speaks for scattering to be dominated by particles of circumference roughly larger than the observing wavelength. 
Smaller grains gradually lose this forward-peaked scattering property and scatter incident light more and more isotropic at all scattering angles \citep[][]{Min2016} as Rayleigh-scattering starts to dominate over Mie-scattering. Indeed, if the dust size distribution is extended to values of $a_{\rm min} \lesssim 0.2\,\mu$m, the contrast between the two sides is continuously being decreased in our synthetic image. This could speak for an evolved dust population (truncated below 0.2$\,\mu$m) within the bright rings. To constrain this further, it has to be seen whether changing the assumption of spherical, compact grains allows the distribution to extend to smaller sizes.

{
\subsubsection{Model shortcomings and degeneracy}}
{Direct comparison of the synthetic images with the real observations shows that in NIR scattered light the outer gap between the rings R$_2$ and R$_3$ is overproduced in width and depth. Additionally, both rings appear brighter in the simulated ALMA image than in the observational counterpart.
We, therefore, warn against taking the results as an unequivocal confirmation of the presence of protoplanets or interpreting the implemented planetary masses as exact predictions. Our simplified hydrodynamical model neglects effects such as self-gravity, dust evolution and dynamics dependent on the vertical dimension. Therefore, it only has the right to aim at a qualitative comparison between observations and our hydrodynamical estimate.
Qualitatively, our model reproduces most features of the observations. Still, the chosen setup is not expected to be unique.} Dedicated studies showed that besides a planet's mass, the width of a gap depends on the level of viscosity and the aspect ratio of the disc \citep[e.g.][]{Crida2006,Kanagawa2016}. Modifying the planetary mass and adapting the other parameters can lead to similar structures as presented. For the inner planet we implemented a mass of 3.7$\,m_{\rm J}$. If responsible for the inner cavity, the lower limit of the planetary mass is given by its capability to efficiently filter out the dust grains that are of sizes large enough to significantly contribute to thermal emission at $1.25\,$mm. This process depends on the companion's mass ratio to the central objects, but again also on disc viscosity and aspect ratio \citep[][]{Weber2018,Haugbolle2019}. It is, therefore, important to discuss the present constraints on those two parameters of the disc environment.
\subsubsection{Implications of binary shadows}
Both stars are assumed to have roughly solar spatial dimensions. This means binary self-occultation only occurs for locations in the disc's atmosphere where, $z/r\lesssim R_\odot /a_{\rm bin}\approx 0.11$. That the shadows are detectable puts a constraint on the scattering surface, $z_{\tau=1}$: it has to be smaller than the value $0.11\times r$ wherever the binary shadows are present. The profile for $z_{\tau=1}$ depends on both the aspect ratio and suface density profile of small dust grains in the disc. At locations where the density of small grains is high, such as in the observed dust rings, the scattering surface can be well above the scale height of the ring as shown by \citet{Stolker2016}. In their specific analysis of the system HD~100546 they find the scattering surface to be roughly three times higher than the disc's pressure scale height. In our model, the aspect ratio at the the inner ring's (R$_2$) location is $h(r=15\AU)\approx 0.037$ which is roughly one third of the demanded occultation condition. For values significantly larger than the implemented scale height the binary shadows disappear from the generated NIR scattered light images. 

\subsubsection{Dust Trapping, Settling and Diffusion}
Ring R$_2$ offers an excellent site to study the size-dependent effect of dust trapping in a radial gas pressure bump. As shown in Fig.~\ref{fig:Hydro results} the smallest dust grains are sufficiently coupled to trace the same Gaussian profile as the gas, which suggests that the NIR scattered light image offers a good reflection of the gas density profile. On the other hand, larger sizes get radially confined towards the centre of the pressure maximum. This is consistent with the observed thin ring in the ALMA band 6 observation. The value of the relative width $\sigma_{\rm d}/\sigma_{\rm g}=0.35$ for grains of $a=178\,\mu$m is in good agreement with the ring width ratio of $\sigma_{\rm ALMA}/\sigma_{\rm VLT}=2\AU/5.3\AU=0.38$ measured in \citetalias{Martinez2021}.

To reproduce the SED at longer wavelengths we had to reduce the vertical spread of decoupled dust grains. We achieve this by assuming the level of vertical dust diffusivity to be diminished by a factor of five with respect to its radial counterpart used in the two-dimensional hydrodynamical simulation. The increased dust settling towards the mid-plane is consistent with a very flared disc and an enhanced dust-to-gas ratio \citep[][]{Lin2019}, as expected in the mid-plane area of the pressure maximum. 
Also in the context of turbulence generated by the magnetorotational instability the vertical component is expected to be systematically smaller than in the radial direction \citep[][]{Zhu2015b}.
In a recent study of the effect of the disc's self-gravity on the diffusion of dust grains, \citet[][]{Baehr2021} found that the diffusivity can be strongly reduced under such consideration, further highlighting a possible difference of radial and vertical dust diffusion behaviour.

{The dust trapping at L$_4$ and L$_5$ does not leave a visible signature in our synthetic image. The existence of these accumulations should depend on the present level of viscosity and the time of evolution of the system as shown in both \citet[][]{Montesinos2020} and \citet[][]{Rodenkirch2021}. Our simulation only covers an evolutionary time of $\sim100\,$kyr and might, therefore, overestimate these dust masses if a possible real planetary system has been evolving for longer.}

{
\subsubsection{Gas kinematics}
The association of the presence of protoplanets with the molecular line emission exposing local deviations from Keplerian rotation puts a certain interest on the rotational profile of putatively planet-forming discs. In general, there are two different types of signatures: {\it(a)} a radially and azimuthally localised transition between sub- and super-Keplerian velocities at the planet's location that can result in a {\it kink} observed in velocity channel maps (as predicted by \citealp{Perez2015} and first observed by \citealp{Pinte2018}) or as a {\it Doppler-flip} \citep[][]{Casassus2019}, and {\it(b)} azimuthally global, symmetric rings of sub-Keplerian rotation inside, and super-Keplerian rotation outside a giant planet's orbit as discussed for and putatively discovered in HD~163296 \citep[][]{Teague2018}. In Fig.~\ref{fig:kinematics} we show the azimuthal velocity profile of our model as deviation from the Keplerian profile, $\Delta v_\varphi = v_\varphi - v_{\rm K}$. 
\begin{figure*}
    \centering
    \includegraphics{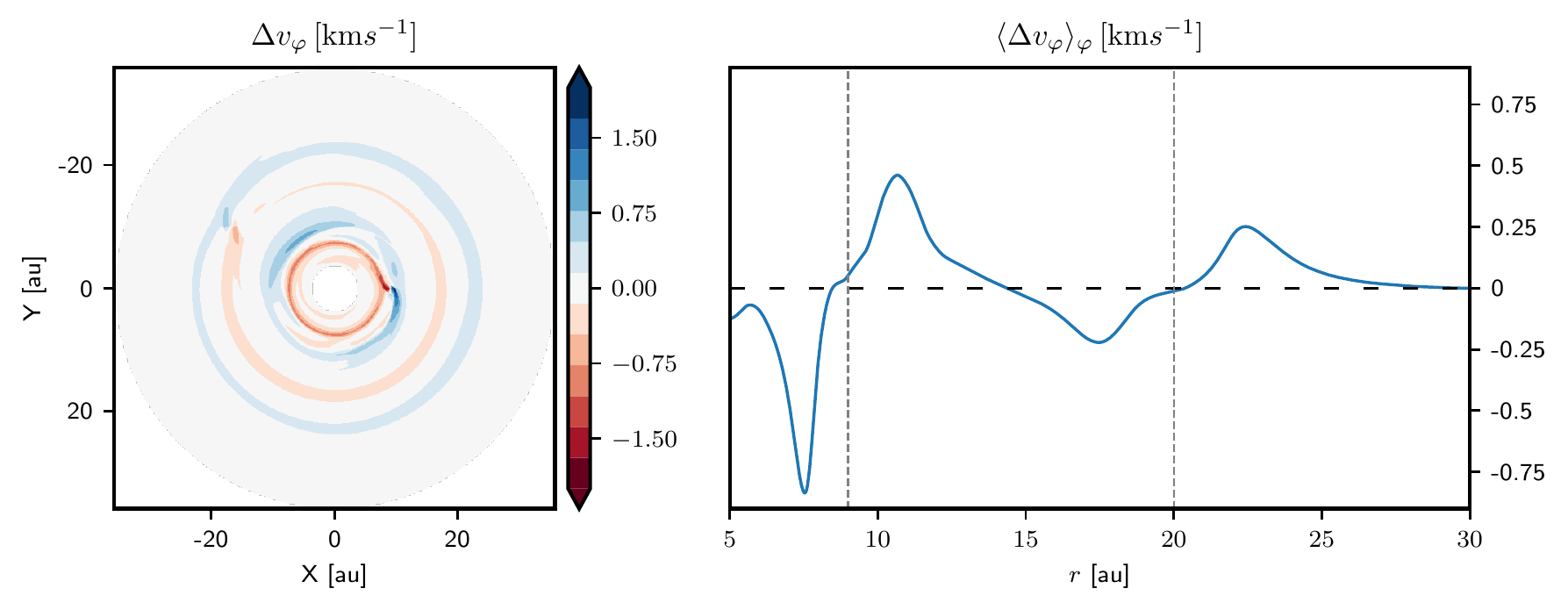}
    \caption{Azimuthal gas velocity as deviation from the local Keplerian velocity. The left panel shows the full two-dimensional disc, the right panel shows the radial profile, as averaged over the azimuthal coordinate.}
    \label{fig:kinematics}
\end{figure*}
One can see that at the location of the outer planet ($v_{\rm K}$ at $20\,{\rm au} =8.8\,{\rm km}\,{\rm s}^{-1}$) the deviation from the Keplerian profile is rather shallow and peaks at only about $3\%$ of the local speed of rotation. While the perturbation cast by the inner planet is more significant (flip of about $10\%$ of local Keplerian velocity, $13.3\,{\rm km}\,{\rm s}^{-1}$), its small separation to the inner stars make it a challenging observational target. Both candidates would demand an extensive observation with very high spectral resolution (to achieve precision in velocity channels), as well as very high spatial resolution. To propose sophisticated kinematical predictions for this system (including synthetic gas observations) three-dimensional simulations are needed. The velocity deviation in Fig.~\ref{fig:kinematics} is representative for the mid-plane. Molecular line emission, however, is typically associated with upper layers of the disc's atmosphere \citep[as e.g. comprehensively explained in][]{Teague2019}, where the planetary effects might be less compulsive. Additionally, the vertical velocity component has to be taken into account for a full kinematical prediction.}

\subsection{Prospect of observing a circumplanetary disc shadow in V4046 Sgr}
\subsubsection{Analysis of the azimuthal intensity profile of scattered light on the inner ring}\label{subsec:analysis}
A close inspection of the SPHERE/IRDIS observation in the upper panel of Fig.~\ref{fig:VLT} reveals the appearance of a third intensity diminution on the bright ring (R$_2$) in the southern, far side of the disc. While the intensity decrements in the eastern and western side of the disc were attributed as effects of binary self-shadowing \citep[][]{DOrazi2019}, this additional indentation has been unaddressed so far.
There are no visible hints for artifacts from spider diffraction spikes in the SPHERE/IRDIS image disfavouring the idea that the detected feature is associated with the instrumentation. 

In Fig.~\ref{fig:az-cut} we show the intensity traced along R$_2$ in a deprojected azimuthal cut at a stellocentric distance where the ring reaches its maximum in scattered light ($r\approx15\AU$). The left panel shows that the dip at the far side is the most prominent localised feature when compared against the background intensity. We fit the data in the shaded area of the left panel by a second-order polynomial (accounting for the background intensity) superimposed by a Gaussian (accounting for the local decrement). The right panel of Fig.~\ref{fig:az-cut} shows the data (every fifth data point), the contribution of the background fit and total fit within the fitted zone. The Gaussian part of the fit is shown in the inlay of the right panel and has a depth of $(\Delta I / I_0)_{\rm min} = -14.9\,$per cent and a standard deviation of $\sigma_{\rm shadow} = 0.115\,$rad, corresponding to a full width at half maximum (FWHM) of 0.27$\,$rad ($15.5^\circ$). We estimate the FWHM of the point spread function (PSF) of the VLT observation to be 0.05$\,$arcsec, which is the diffraction limit of the 8.2$\,$m VLT telescope. This width is constant in the plane of the sky, to calculate the corresponding width in the plane of the disc, one has to take the disc's inclination, $i$, as well as the signals location within the disc, $\hat{r},\hat{\varphi}$, into account:
\begin{equation}
\Delta \varphi_{\rm PSF}=0.05\,{\rm arcsec}\frac{d}{\hat{r}\sqrt{\cos^2i\cos^2\hat{\varphi}+\sin^2\hat{\varphi}}}\,.
\end{equation}
\begin{figure}
    \centering
    \includegraphics[width=\columnwidth]{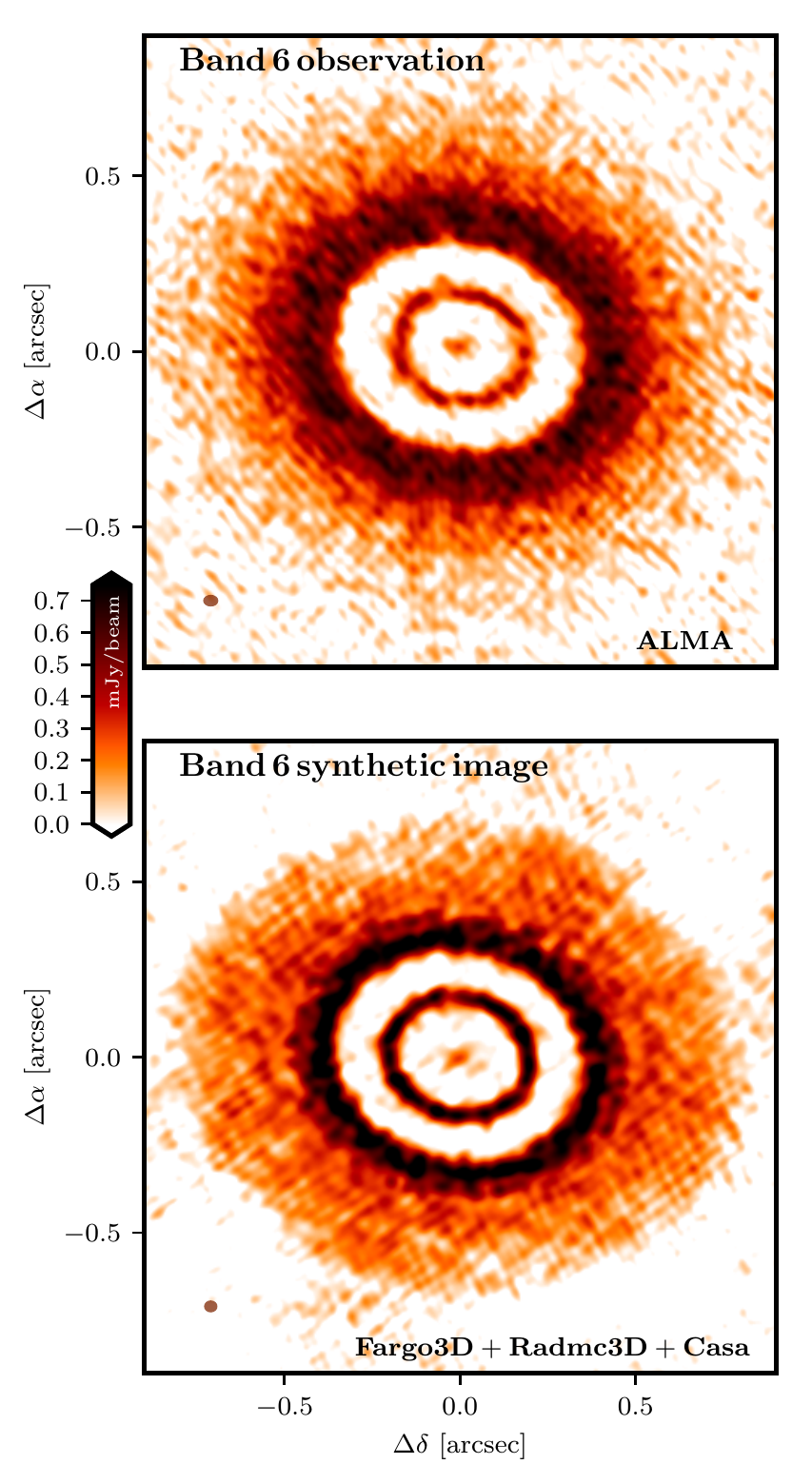}
    \caption{\textit{Top panel:} ALMA band 6 observation reconstructed using the tclean task with Briggs weighting (robust=0). \textit{Bottom Panel:} Adapted synthetic image (at $\lambda_{\rm obs}=1.25\,$mm) including hydro (FARGO3D), RT (RADMC-3D) and image post-processing (CASA). The beam is shown in the bottom left corner for both cases; in both images the intensities are given per beam size \textit{of the upper panel} to make them directly comparable.}
    \label{fig:ALMA}
\end{figure}
The measured location of the signal\footnote{We define the azimuth as zero at the semi-major axis east of north and increasing in clockwise direction.} ($\hat{r}\approx 15\AU$, $\hat{\varphi}\approx 262.5^{\circ}$) is close to the disc's semi-minor axis, where the projection effect is smallest and $\Delta \varphi_{\rm PSF} \approx 0.24\,$ rad. 
The shaded area in the inlay in the right panel of Fig.~\ref{fig:az-cut} shows this azimuthal width associated with the FWHM of the PSF. The inlay shows that the observed signal \textit{is of similar width as the PSF} -- or in other words: the decrement is potentially smoothed by the observational limitations.
\begin{figure*}
    \centering
    \includegraphics[width=\textwidth]{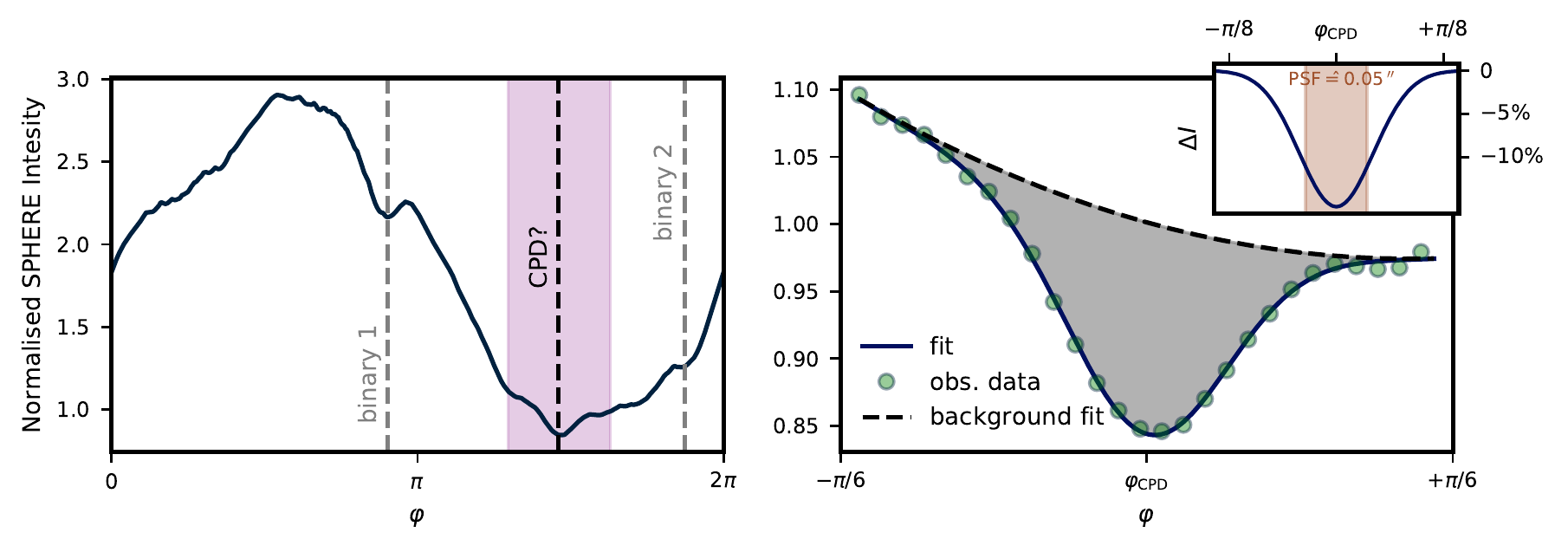}
    \vspace{-0.7cm}
    \caption{Normalised deprojected azimuthal intensity profile of the $Q_\varphi$ data shown in the central panel of Fig.~\ref{fig:VLT}. To trace the azimuthal profile within the ring, we select the distance to the centre where the intensity reaches its maximum. The left panel shows the distribution over the full azimuth, where we highlight the decrements associated with the binary's self-shadowing \citep{DOrazi2019} and the CPD-shadow candidate. We show the selected area around this minimum in the right panel (we display every fifth data point as a circle), centred on the azimuthal location of the dip. The data is fitted by a superposition of a second-order polynomial and a Gaussian. This polynomial contribution accounts for the variable background and is shown in dashed lines. The inlay shows the Gaussian contribution to the fit, where the shaded area contains the estimated point spread function of the observation, FWHM$\approx0.05\,{\rm arcsec}$. Projected into the disc plane and at the radius of the ring, this yields a resolution width of $\Delta \varphi_{\rm R_2}\approx17.6^\circ$.}
    \label{fig:az-cut}
\end{figure*}

The intriguing presence of a third decrement in the PDI data from 13$^{\rm th}$ of March 2016, additional to the binary shadows, automatically presses the question of its origin. Similar intensity diminutions have previously been linked to self-shadowing of central stars \citep[][]{DOrazi2019} or a shadow cast by a tilted inner disc \citep{Marino2015}. However, these possibilities typically demand the presence of a similar feature on the opposing side of the shadowed ring. Such a counterpart cannot be detected in the azimuthal profile. Therefore, we propose an alternative scenario: the indentation may be linked to the presence of a circumplanetary disc (CPD) around a giant planet embedded in the inner cavity. In this scenario a direct detection of this candidate would be impeded by the coronagraph. 

The third decrement (as well as the binary shadows) is smoothed by the PSF, and its profile is approximately Gaussian in shape. We can use the depth of the decrement to constrain the size of the CPD.
The important scale in describing the size of a CPD is the Hill-radius, i.e. the maximum distance around the planet in which an object can remain on a bound orbit:
\begin{equation}\label{equ:Hill}
    R_{\rm Hill} = r_{\rm P} \left(\frac{m_{\rm P}}{3 m_{\rm b}} \right)^{1/3}\,.
\end{equation}
The radial extend of the CPD is typically assumed to be a fraction of this scale, $R_{\rm CPD} = \Lambda R_{\rm Hill}$, with $\Lambda<1$. Hence, the azimuthal angle covered by the CPD is given by $\Delta \varphi_{\rm CPD} \approx d_{\rm CPD}/r_{\rm P}$, where $d_{\rm CPD}=2R_{\rm CPD}$ is the CPD's diameter. If the CPD is a sharp, optically thick body, creating an absolute shadow in the form of a step function, its size can be related to the intensity diminution:
\begin{equation}\label{equ:deltaCPD}
    \Delta \varphi_{\rm CPD} = \frac{\int\Delta I {\rm d}\varphi}{I_{0}} = \sqrt{2\pi}\sigma_{\rm shadow} \left(\frac{|\Delta I|}{I_0}\right)_{\rm min} \,.
\end{equation}
  We can thus express the CPD size as a function of the observed signal and dependent on the planet's orbital distance:
\begin{equation}\label{equ:R-CPD}
    R_{\rm CPD} = \frac{r_{\rm P}\sqrt{2\pi}\sigma_{\rm shadow}}{2} \left(\frac{|\Delta I|}{I_0}\right)_{\rm min}\,.
\end{equation}
We use the link of $R_{\rm CPD}$ to the Hill radius and its definition. This allows to estimate a planetary mass, independent of its orbital distance:
\begin{equation}\label{equ:planetmass}
    m_{\rm P} =3m_{\rm b}\left[\frac{\sqrt{2\pi}\sigma_{\rm shadow}}{2\Lambda} \left(\frac{|\Delta I|}{I_0}\right)_{\rm min}\right]^{3}\,.
\end{equation}
The Gaussian fit in the right panel of Fig.~\ref{fig:az-cut} shows a minimum value of $(|\Delta I|/I_0)_{\rm min} = 0.149$. Assuming that $\Lambda = 0.4$ the planetary host would be of $\approx 0.85\,{\rm M}_{\rm J}$. This constitutes the lowest possible planetary mass, as it assumes that the unresolved shadow is total and in the shape of a step-function. 

\subsubsection{Comparison to simulations}\label{subsec:simulations}
To test the hypothesis of a shadow cast by circumplanetary material, we include an artificial CPD within the radiative transfer calculations described in Section~\ref{subsec:RTmodel}. The reason for including this disc as a parametric model is two-fold: \textit{(a)} the formation and evolution of CPDs is in itself a complex and degenerate topic that depends on the adequate knowledge of input parameters and treatment of sensitive processes (see references in the introduction of \citealp{Zhu2015a}, as well as \citealp{Fung2019} for more recent work with a focus on the treatment of thermodynamics) 
and \textit{(b)} it capacitates us to easily control geometrical and physical disc parameters and compare the different effects on the proposed shadow.

For the model of the implemented CPD we consider a radial extend of $R_{\rm CPD} =0.4R_{\rm Hill}$, consistent with dedicated CPD studies \citep[e.g.][]{D'Angelo2003,Ayliffe2009}.
We define the radial distance from any point in the CPD's mid-plane to the planet as $\Tilde{R} = |\mathbf{r} - \mathbf{r}_{\rm p}|$, and power-laws for temperature and densities\footnote{to clearly distinguish quantities in the planet-centred frame from quantities in the stellocentric frame, we henceforth use the tilde as a symbol for those corresponding to the CPD.}:
\begin{equation}\label{equ:powelaw-CPD}
     \Tilde{\Sigma} = \Tilde{\Sigma}_{\rm 0} \left(\frac{\Tilde{R}}{{R}_{\rm CPD}}\right)^{-3/4}\,. 
\end{equation}
The power-law exponent is taken from an estimate by \citet{Ayliffe2009} and equation~(\ref{equ:powelaw-CPD}) is applied within $0.02R_{\rm Hill}<\Tilde{R}<R_{\rm CPD}$. 
We then implement a CPD corresponding to a massive planet, $m_{\rm P_1} = 3.7\,{\rm M}_{\rm J}$ at $r_{\rm P_1} = 9\AU$, and set the mass of solids to $m_{\rm CPD} = 2\times10^{-4}m_{\rm P_1}$, motivated by the mass ratio between the Jovian moons and Jupiter. Further, the grain size distribution and dust-to-gas ratio in the CPD are assumed to be inherited from the PPD. 
We use the same RT model as described in Section~\ref{sec:obs} but to adequately model the comparably small CPD inside the RT grid and avoid spending a lot of computational power on empty cells, we make use of RADMC-3D's built-in adaptive mesh refinement capability (option: Layer-style AMR grid). 
We select a zone, centred on the location of the planet, with an edge length of $R_{\rm Hill}$ in radial and azimuthal direction and spanning over the entire polar angle. 
Within, we increase the resolution by a factor of $2^4$ in each spatial direction. 

To cast a perceivable shadow, sufficient dust particles have to be high above the PPD's mid-plane at the location of the CPD. We thus assume the CPD to be very vertically inflated. In Appendix~\ref{appendix:inclined} we discuss an alternative possibility of a CPD inclined with respect to the outer PPD. 
We set a constant aspect ratio of $\Tilde{h} = H_{\rm g}/\Tilde{R} = 0.5$. {Assuming vertical hydrostatical equilibrium dominated by thermal pressure and gravity, this aspect ratio can be translated into a temperature of $\sim 600\,{\rm K}$ at the CPD's outer edge, which is roughly consistent with models presented in \citet[][]{Szulagyi2017b} for a giant planet formed by core accretion (in contrast to gravitational instability)}. The CPD is assumed to be co-planar with the PPD. {We want to highlight that our implemented CPD model, defined by power-law profiles, differs from the result in \citet[][]{Montesinos2021}, where a strongly inflated inner edge of the CPD is responsible for shadowing effects. The numerical and theoretical motivation of the model implemented here can be found in several dedicated studies \citep[e.g.][]{D'Angelo2003,Ayliffe2009,Szulagyi2017,Schulik2019}. The shadowing scenario for a CPD as proposed by \citet[][]{Montesinos2021} is already extensively discussed in their work.}
\begin{figure}
    \centering
    \includegraphics[width=\columnwidth]{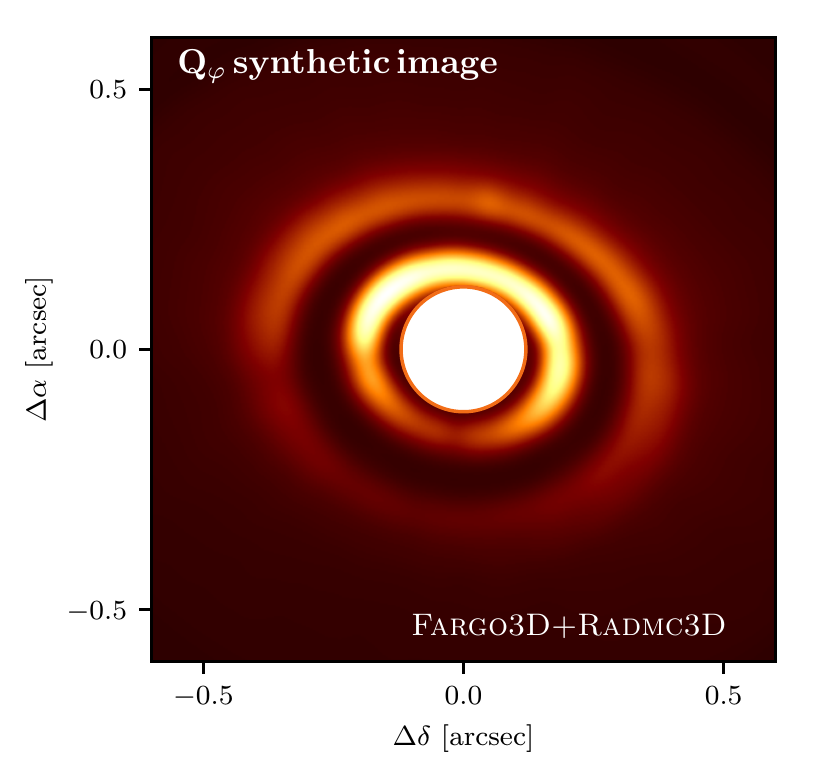}
    \vspace{-0.5cm}
    \caption{Same as bottom panel of Fig.~\ref{fig:VLT} but with a vertically inflated CPD included at the southern, far side of the image as described in Sec.~\ref{subsec:simulations}}
    \label{fig:CPDinflated}
\end{figure}
The resulting synthetic VLT/SPHERE image can be seen in Fig.~\ref{fig:CPDinflated}. The global image is equal to the bottom panel in Fig.~\ref{fig:VLT} but also reproduces the intensity decrement at the southern side due to the shadow cast by the implemented CPD. 
\begin{figure}
    \centering
    \includegraphics[width=\columnwidth]{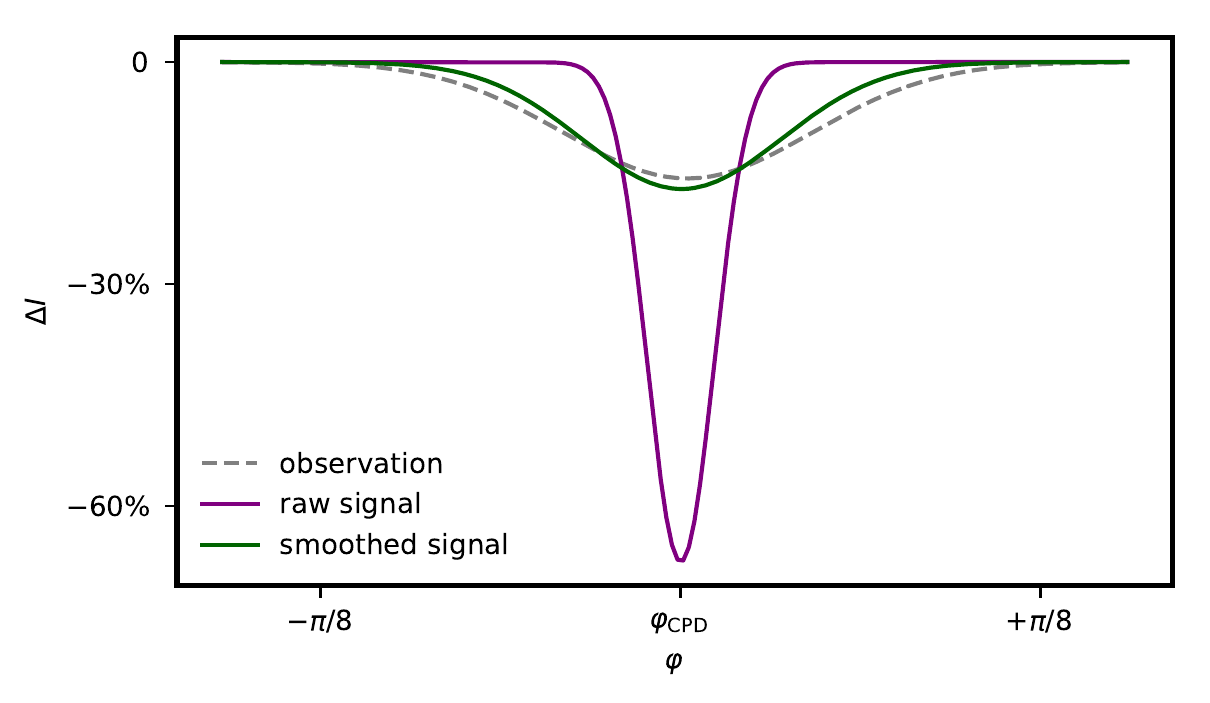}
    \vspace{-0.5cm}
    \caption{Intensity decrement after background subtraction for planet of $3.7\,m_{\rm J}$ at $9\AU$: The dashed grey line shows the fit to the SPHERE/IRDIS data as shown in Fig.~\ref{fig:az-cut}, the purple line shows the raw signal of a simulated CPD shadow as described in Sec.~\ref{subsec:simulations}, the green line shows the same signal when smoothed by a Gaussian to account for the PSF limitation of the observation.}
    \label{fig:az-cutsimnew}
\end{figure}
Fig.~\ref{fig:CPDinflated} shows that the CPD structure embedded in our RT model casts an observable shadow on the southern side of the disc. To compare this shadow more quantitatively to the observation, we perform the same extraction as we did to the real data as was presented in Fig.~\ref{fig:az-cut}: produce an azimuthal profile of the bright ring and combine a second-order polynomial with a Gaussian to fit the data.
One advantage of the simulated data is that we are not limited by the PSF of the telescope. In the synthetic images shown in the right panel of Fig.~\ref{fig:VLT} and in Fig.~\ref{fig:CPDinflated} we convolved the intensity field with the expected PSF to make the simulation directly comparable to the observation. In Fig.~\ref{fig:az-cutsimnew} we additionally regard the azimuthal intensity profile prior to convolution which reveals the physical shadow cast by the simulated CPD. 

\paragraph*{Close-in Planet}
So far, we only considered the scenario that the decrement observed on R$_2$ stems from a planet located at an orbital radius of $9\AU$ from the binary, motivated by the hydrodynamical predictions. We showed by RT modelling that a certain CPD model around a giant protoplanet can explain the observed intensity decrement. 
At the location of $r_{\rm P}=9\AU$, the shadow-casting object is sufficiently distanced that the spatial separation of the light sources does not notably impact the shadowing. This changes if the spatial extent of the shadowing object is more similar to the light source separation. A geometrical analysis of a scenario where the light source separation becomes relevant is presented in Appendix~\ref{appendix:binary}.
Here, we want to discuss the idea that, in principle, the diminution could originate from an additional giant planet within the cavity, closer to the central binary. 
We set up a CPD at a distance to the star of $r=1\AU$, to investigate this scenario.
We use the same CPD setup as previously. As its Hill radius scales linearly with the planet's orbital distance around the binary, the CPD is significantly smaller than in the previous case. The solid angle that it covers, however, remains the same.
\begin{figure}
    \centering
    \includegraphics[width=\columnwidth]{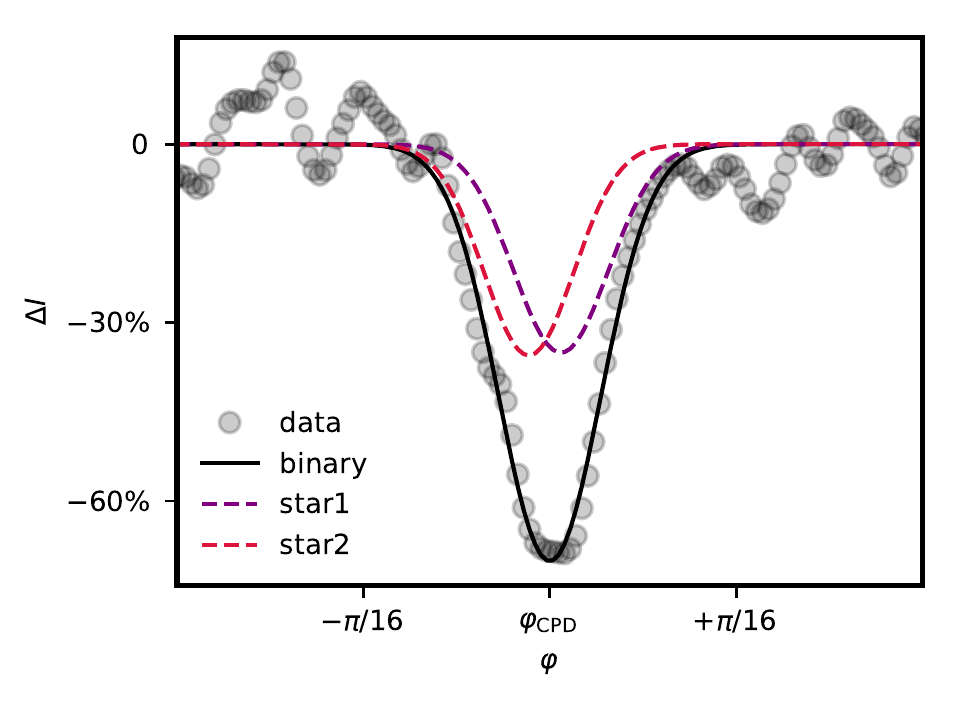}
    \vspace{-0.7cm}
    \caption{Similar as Fig.~\ref{fig:az-cutsimnew}: unconvolved intensity decrement after background subtraction, for a CPD around a planet of $m_{\rm P_1} = 3.7\,m_{\rm J}$ at $r_{\rm P}=1\AU$. The dashed lines show the cases corresponding to the contribution of each individual binary component. The solid line shows the fit for the case of including both stars. The circles show the fitted data points.}
    \label{fig:close-in}
\end{figure}
Fig.~\ref{fig:close-in} shows the resulting intensity decrease due to shadowing. We show the unconvolved signals, highlighting the contributions of the individual binary components. These signals were calculated by consecutively performing the RT with one of the stars turned off {and fitting a Gaussian to the correspondingly cast CPD shadow. The circles show the data points around the azimuthal location of the CPD shadow in the case of both binary components included. The oscillations outside the Gaussian area are due to noise from the radiative transfer and can be reduced by increasing the number of photon packages in the ray-tracing procedure.} 
The comparison of the Gaussians shows well the magnification of the signal due to the superposition of two individual shadows {as discussed in Appendix~\ref{appendix:binary}}. 

A lower limit for the orbit of the shadow casting planet can be obtained by considering an inner region of orbital instability around the binary. At close enough distances the planet's mean-motion resonances will intersect with those of the binary which can stimulate the planetary eccentricity and eventually induce the planet's ejection from the system or collision with one of the binary constituents \citep[see][ and references therein]{Sutherland2019}. We estimate the critical orbital distance, $r_{\rm c}$, according to equation~(3) presented in \citet{Holman1999}, where we drop contributions of the binary's eccentricity and write the expression in terms of the nowadays more commonly used binary mass fraction, $q_{\rm b} = m_{\star,2}/m_{\star,1}\approx0.94$:
\begin{equation}
    \frac{r_{\rm c}}{a_{\rm b}} \approx 1.6 + 4.12\frac{q_{\rm b}}{q_{\rm b}+1} - 5.01\left(\frac{q_{\rm b}}{q_{\rm b}+1}\right)^2 = 2.42\,.
\end{equation}
This value is also in good agreement with the prediction by \citet{Dvorak1989} from numerical tests and translates into a lower limit for the planet candidate of $r_{\rm c} = 0.1\AU$. The stability limit is also present in the detected exoplanet population in circumbinary systems \citep[][we also point to this reference for the assessment of where our predicted orbits fall in detected binary exoplanets statistics]{Martin2018}.
The (admittedly limited) data of detected planets around close-in binaries presented in Fig. 4 of \citet{Martin2018} shows that the critical radius is not only a lower boundary but also a location of accumulation of detection. This phenomenon is typically interpreted as a consequence of planetary inward migration and subsequent stoppage at the truncated inner cavity created by the binary {as studied for specific Kepler systems by \citet[][]{Pierens2013} and \citet[][]{Kley2014}, and recently in a more generalised approach by \citet[][]{Penzlin2021}.}

\subsubsection{ALMA observability}
In the existing ALMA observation recorded in 2017 \citepalias{Martinez2021} no significant emission is discernible from within the cavity except the emission confined to the very centre. At the proposed orbital distance of $r_{\rm P}=9\AU$, a CPD around a hot protoplanet that contains grains of up to mm-sizes should be a visible feature \citep[][]{Szulagyi2018}, even considering the comparably short total integration time of only about $\tau_{\rm int}\approx13\,$min on target. This does not rule out the existence of a CPD within the cavity, it could also indicate that the assumptions of a high temperature and large grains is not accurate for this case.
\begin{figure}
    \centering
    \includegraphics[width=\columnwidth]{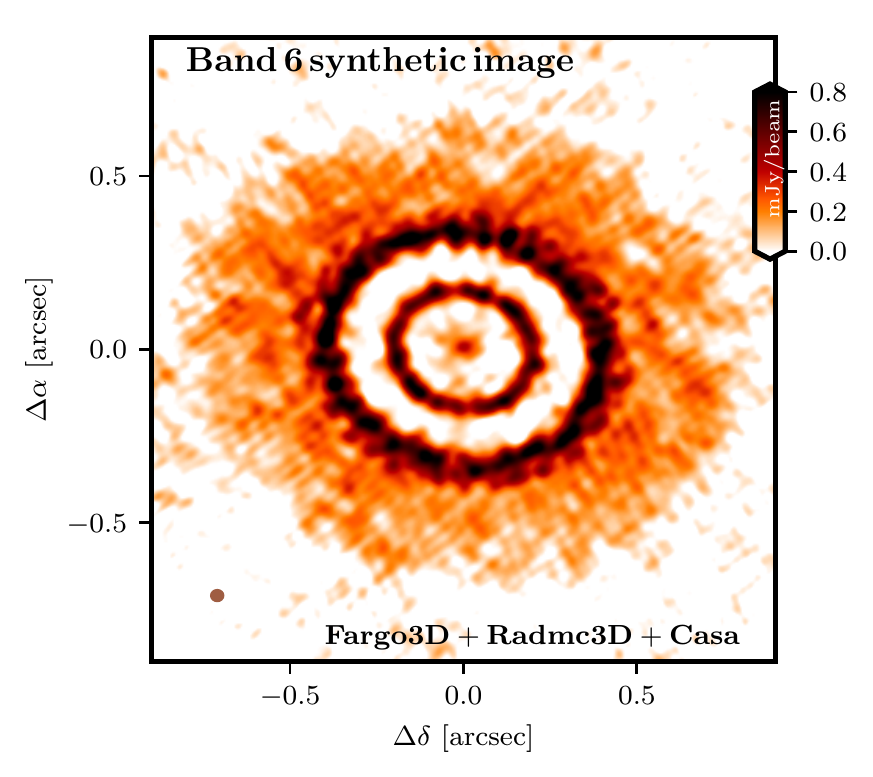}
    \vspace{-0.3cm}
    \caption{Same as bottom panel of Fig.~\ref{fig:ALMA} but with CPD model included at the southern extreme in the inner cavity at $r=9\AU$.}
    \label{fig:ALMA-CPD}
\end{figure}
Although \citet[][]{Szulagyi2021} found that in three-dimensional simulations even mm-grains follow vertical accretion streams onto the CPD, their presented simulations of 100--200 orbits do not allow the gap to fully develop and, therefore, might underestimate the filtering efficiency.
In the two-dimensional hydrodynamical simulation presented in Fig.~\ref{fig:Hydro results}, we find that the dust species representing larger sizes are prohibited from entering the gap area \citep[consistent with][]{Weber2018}. Therefore, we assume for our CPD model that the dust size distribution is truncated at $a_{\rm max}\approx 10\,\mu$m.
A further important quantity for the visibility of the CPD is its temperature. Similar to \citet{Benisty2021} we assume the CPD temperature to be dominated by the PPD environment at its outer edge and set the temperature there to $\Tilde{T}_{\rm CPD}(\Tilde{R} = R_{\rm CPD}) = 42\,$K. We further implement a radial temperature profile within the CPD of $\Tilde{T} \propto \Tilde{R}^{-0.7}$ \citep[motivated by][]{Ayliffe2009}.\\
For a setup similar to the existing ALMA band~6 observation, this CPD model emits millimetre fluxes below the level of noise (see Fig.~\ref{fig:ALMA-CPD}) -- while still reproducing the shadow feature in the NIR. However, increasing the integration time reveals a detectable intensity feature above noise level at the CPD's location.
\begin{figure}
    \centering
    \includegraphics[width=\columnwidth]{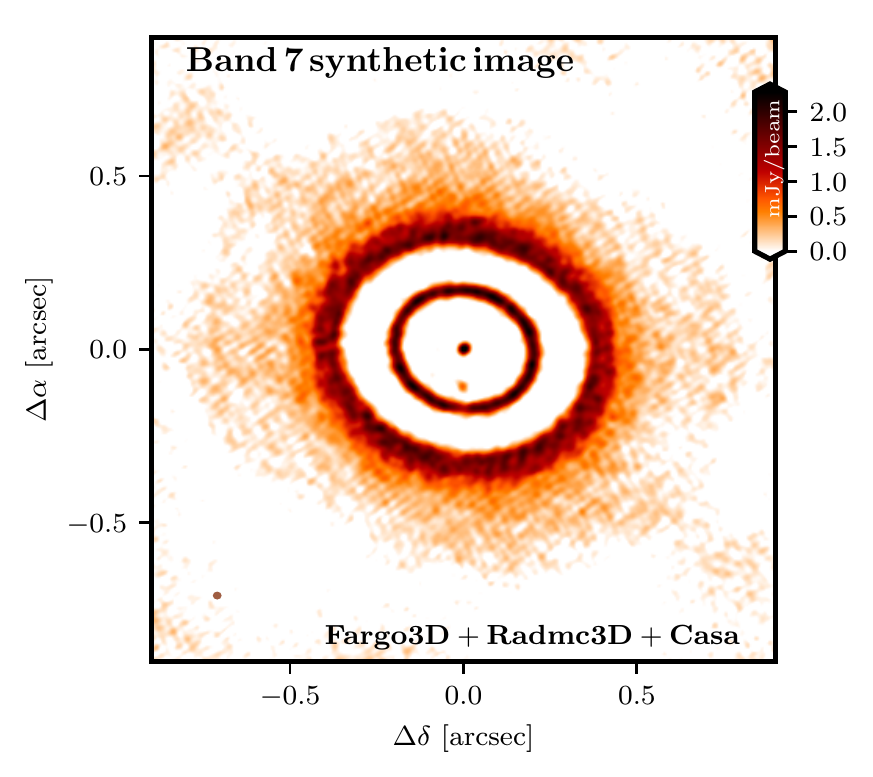}
    \vspace{-0.5cm}
    \caption{Same as Fig.~\ref{fig:ALMA-CPD} but observed in ALMA band 7 ($\lambda_{\rm obs} = 87\,\mu$m) and integration time increased to $\tau_{\rm int}=100\,$min.}
    \label{fig:ALMA-longint}
\end{figure}
Observing the disc at shorter wavelengths additionally increases the CPD signal. Fig~\ref{fig:ALMA-longint} shows an ALMA band 7 ($\lambda_{\rm obs}=870\,\mu$m) synthetic image with increased integration time of $\tau_{\rm int} =100\,$min. The CPD feature is clearly detectable in this setup.

For a close-in CPD, the constraints on the emitted flux are less stringent. Primarily, a CPD closer to the star displays a smaller emitting surface. In addition, the central emission visible in the ALMA observation allows for the presence of a visible CPD, within an orbit of $\lesssim 1\AU$.

In some extreme cases of a large and heavy CPD within the inner cavity, the synthetic ALMA image can also show reduced emission at the location of the NIR shadow of the CPD. We warn against taking this as a valid prediction. The thermal balance in the RT calculations do not include the advection of internal energy, nor any dynamical effect; they regard a scenario of a static shadow on static disc material. In reality, however, the shadow is moving azimuthally according to the Keplerian velocity at the CPD's location and the dust in R$_{2}$ is also in (slower) Keplerian rotation. This results in a net azimuthal motion of the shadow over the material in the ring. The approximation of a static disc is valid when considering the shadow in scattered light, as here the signal is transmitted immediately when the diminution of incident light occurs. In the mm-regime, however, the decrement is due to a temperature drop at the mid-plane within the shadowed area, resulting in reduced thermal emission from the dust grains.
Further, the disk thermal radiation acts as an extended light source which, if optically thin in the Rosseland sense, will smooth out any decrement.
Consequently, \cite{Casassus2019b} showed that this temperature drop can be dramatically reduced in the dynamical case and depends on the shadow's crossing time and local cooling time-scales.\\

The number of exoplanets found around close binary stars (with an orbital period of just a few days) is reduced compared to other systems \citep[][]{Martin2018}. {By taking into account binary systems that were previously thought to be single stars, \citet{Bonavita2020} recently confirmed that while no significant difference of planet occurence can be detected around binaries in genereal, very close binaries seem to conspiciously lack planetary companions. The detection of forming planets at such sites} -- as suggested for V4046~Sgr from our hydrodynamical comparison -- is therefore of increased interest.\\
The detected indentation of NIR scattered light on the bright ring of V4046$\,$Sgr is without counterpart on the opposing side of the disc. This rules out some of the mechanisms previously associated with shadowing (e.g. by an inner warped disc, \citealp{Marino2015}, or self-shadowing of two luminous bodies, \citealp{DOrazi2019}).
Yet, with the data at hand, it is not possible to indisputably link the observed intensity decrement to a planetary origin. 
Future observations hold the promise to confirm a recurring intensity indentation and facilitate the extraction of orbital elements under the assumption that the decrement is the shadow of a Keplerian object.

\section{Conclusions}
Employing hydrodynamical and radiative transfer simulations, we were able to recover the observed structure of the circumbinary disc V4046~Sgr. 
\begin{itemize}
    \item The radial density variability in the outer disc is consistent with the presence of two giant planets, one at $\approx 9\AU$ of a few Jupiter masses, and one at $\approx20\AU$ of roughly Jupiter mass.
    \item In our hydrodynamical multifluid simulation a radial pressure maximum is created between the two planets at roughly 15$\AU$, where dust can be trapped. The different radial widths of the ring in VLT and ALMA observations supports a size-dependent balance between headwind-induced confinement and spread by diffusion. This ring offers an excellent site to study a dust trap in multifrequency analysis and consequently infer limits on the level of diffusivity.
    \item Comparable to the two previously addressed decrements associated with the binary's self-shadowing, we report an additional azimuthally-confined intensity decrement on the inner ring found in the VLT/SPHERE polarised scattered light image. This decrement can be reproduced by a shadow cast by a circumplanetary disc around a giant planet in the inner cavity.
    \item We delineated how a detailed analysis of such a shadow can serve to infer a protoplanetary candidate's location and the size of its associated circumplanetary disc.
\end{itemize}

\section*{Acknowledgements}
We thank the anonymous referee for constructive comments and motivating discussion of our presented results. We further thank Rafael Martinez-Brunner for advice on the choice of input parameters to the model of the circumbinary disc.
P.W. acknowledges support from ALMA-ANID postdoctoral fellowship 31180050.
S.C. and S.P. acknowledge support from FONDECYT grants 1211496 and 1191934. 
The hydrodynamical simulation was performed on GPU computing nodes funded with a research grant from the Danish Center for Scientific Computing and are hosted at the University of Copenhagen HPC facility.
This work made use of the Puelche cluster hosted at CIRAS/USACH and financed by Fondecyt grant 1191934.

\section*{Software}
This work has made use of FARGO3D \citep{PBLL2016,PBLL2019} for hydrodynamical simulations, RADMC-3D \citep{Dullemond2012} for radiative transfer calculations and CASA \citep[][]{McMullin2007} for ALMA image synthesis, as well as IPython \citep{ipython}, NumPy \citep{numpy} and Matplotlib \citep{Matplotlib} for data analysis and creating figures.

\section*{Data availability}
The Gemini/GPI data was provided after private communication as a courtesy of V. Rapson. The raw SPHERE/IRDIS data can be downloaded from the ESO archive using project code 096.C-0523(A). The raw ALMA data can be found in the ALMA archive using the project identification 2017.1.01167.S. The IRDAP-reduced version of the SPHERE/IRDIS data and a clean version of the ALMA data will be made available in fits format in the supplementary material of \citet[][subm.]{Martinez2021}.

\bibliographystyle{mnras}
\bibliography{ref}

\begin{thebibliography}{}
\makeatletter
\relax
\def\mn@urlcharsother{\let\do\@makeother \do\$\do\&\do\#\do\^\do\_\do\%\do\~}
\def\mn@doi{\begingroup\mn@urlcharsother \@ifnextchar [ {\mn@doi@}
  {\mn@doi@[]}}
\def\mn@doi@[#1]#2{\def\@tempa{#1}\ifx\@tempa\@empty \href
  {http://dx.doi.org/#2} {doi:#2}\else \href {http://dx.doi.org/#2} {#1}\fi
  \endgroup}
\def\mn@eprint#1#2{\mn@eprint@#1:#2::\@nil}
\def\mn@eprint@arXiv#1{\href {http://arxiv.org/abs/#1} {{\tt arXiv:#1}}}
\def\mn@eprint@dblp#1{\href {http://dblp.uni-trier.de/rec/bibtex/#1.xml}
  {dblp:#1}}
\def\mn@eprint@#1:#2:#3:#4\@nil{\def\@tempa {#1}\def\@tempb {#2}\def\@tempc
  {#3}\ifx \@tempc \@empty \let \@tempc \@tempb \let \@tempb \@tempa \fi \ifx
  \@tempb \@empty \def\@tempb {arXiv}\fi \@ifundefined
  {mn@eprint@\@tempb}{\@tempb:\@tempc}{\expandafter \expandafter \csname
  mn@eprint@\@tempb\endcsname \expandafter{\@tempc}}}

\bibitem[\protect\citeauthoryear{{Andrews} et~al.,}{{Andrews}
  et~al.}{2018}]{Andrews2018}
{Andrews} S.~M.,  et~al., 2018, \mn@doi [\apjl] {10.3847/2041-8213/aaf741},
  \href {https://ui.adsabs.harvard.edu/abs/2018ApJ...869L..41A} {869, L41}

\bibitem[\protect\citeauthoryear{{Avenhaus} et~al.,}{{Avenhaus}
  et~al.}{2018}]{Avenhaus2018}
{Avenhaus} H.,  et~al., 2018, \mn@doi [\apj] {10.3847/1538-4357/aab846}, \href
  {https://ui.adsabs.harvard.edu/abs/2018ApJ...863...44A} {863, 44}

\bibitem[\protect\citeauthoryear{{Ayliffe} \& {Bate}}{{Ayliffe} \&
  {Bate}}{2009}]{Ayliffe2009}
{Ayliffe} B.~A.,  {Bate} M.~R.,  2009, \mn@doi [\mnras]
  {10.1111/j.1365-2966.2009.15002.x}, \href
  {https://ui.adsabs.harvard.edu/abs/2009MNRAS.397..657A} {397, 657}

\bibitem[\protect\citeauthoryear{{Baehr} \& {Zhu}}{{Baehr} \&
  {Zhu}}{2021}]{Baehr2021}
{Baehr} H.,  {Zhu} Z.,  2021, \mn@doi [\apj] {10.3847/1538-4357/abddb4}, \href
  {https://ui.adsabs.harvard.edu/abs/2021ApJ...909..136B} {909, 136}

\bibitem[\protect\citeauthoryear{{Ballabio}, {Nealon}, {Alexander}, {Cuello},
  {Pinte}  \& {Price}}{{Ballabio} et~al.}{2021}]{Ballabio2021}
{Ballabio} G.,  {Nealon} R.,  {Alexander} R.~D.,  {Cuello} N.,  {Pinte} C.,
  {Price} D.~J.,  2021, \mn@doi [\mnras] {10.1093/mnras/stab922}, \href
  {https://ui.adsabs.harvard.edu/abs/2021MNRAS.504..888B} {504, 888}

\bibitem[\protect\citeauthoryear{{Banzatti}, {Pinilla}, {Ricci}, {Pontoppidan},
  {Birnstiel}  \& {Ciesla}}{{Banzatti} et~al.}{2015}]{Banzatti2015}
{Banzatti} A.,  {Pinilla} P.,  {Ricci} L.,  {Pontoppidan} K.~M.,  {Birnstiel}
  T.,   {Ciesla} F.,  2015, \mn@doi [\apjl] {10.1088/2041-8205/815/1/L15},
  \href {https://ui.adsabs.harvard.edu/abs/2015ApJ...815L..15B} {815, L15}

\bibitem[\protect\citeauthoryear{{Bell}, {Cassen}, {Klahr}  \&
  {Henning}}{{Bell} et~al.}{1997}]{Bell1997}
{Bell} K.~R.,  {Cassen} P.~M.,  {Klahr} H.~H.,   {Henning} T.,  1997, \mn@doi
  [\apj] {10.1086/304514}, \href
  {https://ui.adsabs.harvard.edu/abs/1997ApJ...486..372B} {486, 372}

\bibitem[\protect\citeauthoryear{{Benisty} et~al.,}{{Benisty}
  et~al.}{2021}]{Benisty2021}
{Benisty} M.,  et~al., 2021, \mn@doi [\apjl] {10.3847/2041-8213/ac0f83}, \href
  {https://ui.adsabs.harvard.edu/abs/2021ApJ...916L...2B} {916, L2}

\bibitem[\protect\citeauthoryear{{Ben{\'\i}tez-Llambay} \&
  {Masset}}{{Ben{\'\i}tez-Llambay} \& {Masset}}{2016}]{PBLL2016}
{Ben{\'\i}tez-Llambay} P.,  {Masset} F.~S.,  2016, \mn@doi [\apjs]
  {10.3847/0067-0049/223/1/11}, \href
  {https://ui.adsabs.harvard.edu/abs/2016ApJS..223...11B} {223, 11}

\bibitem[\protect\citeauthoryear{{Ben{\'\i}tez-Llambay}, {Krapp}  \&
  {Pessah}}{{Ben{\'\i}tez-Llambay} et~al.}{2019}]{PBLL2019}
{Ben{\'\i}tez-Llambay} P.,  {Krapp} L.,   {Pessah} M.~E.,  2019, \mn@doi
  [\apjs] {10.3847/1538-4365/ab0a0e}, \href
  {https://ui.adsabs.harvard.edu/abs/2019ApJS..241...25B} {241, 25}

\bibitem[\protect\citeauthoryear{{Bertrang}, {Flock}, {Keppler}, {Trifonov},
  {Penzlin}, {Avenhaus}, {Henning}  \& {Montesinos}}{{Bertrang}
  et~al.}{2020}]{Bertrang2020}
{Bertrang} G. H.~M.,  {Flock} M.,  {Keppler} M.,  {Trifonov} T.,  {Penzlin} A.
  B.~T.,  {Avenhaus} H.,  {Henning} T.,   {Montesinos} M.,  2020, arXiv
  e-prints, \href {https://ui.adsabs.harvard.edu/abs/2020arXiv200711565B} {p.
  arXiv:2007.11565}

\bibitem[\protect\citeauthoryear{{Bohren} \& {Huffman}}{{Bohren} \&
  {Huffman}}{1983}]{Bohren1983}
{Bohren} C.~F.,  {Huffman} D.~R.,  1983, {Absorption and scattering of light by
  small particles}

\bibitem[\protect\citeauthoryear{{Bonavita} \& {Desidera}}{{Bonavita} \&
  {Desidera}}{2020}]{Bonavita2020}
{Bonavita} M.,  {Desidera} S.,  2020, \mn@doi [Galaxies]
  {10.3390/galaxies8010016}, \href
  {https://ui.adsabs.harvard.edu/abs/2020Galax...8...16B} {8, 16}

\bibitem[\protect\citeauthoryear{{Brown-Sevilla} et~al.,}{{Brown-Sevilla}
  et~al.}{2021}]{Brown-Sevilla2021}
{Brown-Sevilla} S.~B.,  et~al., 2021, arXiv e-prints, \href
  {https://ui.adsabs.harvard.edu/abs/2021arXiv210713560B} {p. arXiv:2107.13560}

\bibitem[\protect\citeauthoryear{{Byrne}}{{Byrne}}{1986}]{Byrne1986}
{Byrne} P.~B.,  1986, Irish Astronomical Journal, \href
  {https://ui.adsabs.harvard.edu/abs/1986IrAJ...17..294B} {17, 294}

\bibitem[\protect\citeauthoryear{{Casassus} \& {P{\'e}rez}}{{Casassus} \&
  {P{\'e}rez}}{2019}]{Casassus2019}
{Casassus} S.,  {P{\'e}rez} S.,  2019, \mn@doi [\apjl]
  {10.3847/2041-8213/ab4425}, \href
  {https://ui.adsabs.harvard.edu/abs/2019ApJ...883L..41C} {883, L41}

\bibitem[\protect\citeauthoryear{{Casassus}, {P{\'e}rez}, {Osses}  \&
  {Marino}}{{Casassus} et~al.}{2019}]{Casassus2019b}
{Casassus} S.,  {P{\'e}rez} S.,  {Osses} A.,   {Marino} S.,  2019, \mn@doi
  [\mnras] {10.1093/mnrasl/slz059}, \href
  {https://ui.adsabs.harvard.edu/abs/2019MNRAS.486L..58C} {486, L58}

\bibitem[\protect\citeauthoryear{{Crida}, {Morbidelli}  \& {Masset}}{{Crida}
  et~al.}{2006}]{Crida2006}
{Crida} A.,  {Morbidelli} A.,   {Masset} F.,  2006, \mn@doi [\icarus]
  {10.1016/j.icarus.2005.10.007}, \href
  {https://ui.adsabs.harvard.edu/abs/2006Icar..181..587C} {181, 587}

\bibitem[\protect\citeauthoryear{{D'Angelo}, {Henning}  \& {Kley}}{{D'Angelo}
  et~al.}{2003}]{D'Angelo2003}
{D'Angelo} G.,  {Henning} T.,   {Kley} W.,  2003, \mn@doi [\apj]
  {10.1086/379224}, \href
  {https://ui.adsabs.harvard.edu/abs/2003ApJ...599..548D} {599, 548}

\bibitem[\protect\citeauthoryear{{D'Orazi} et~al.,}{{D'Orazi}
  et~al.}{2019}]{DOrazi2019}
{D'Orazi} V.,  et~al., 2019, \mn@doi [Nature Astronomy]
  {10.1038/s41550-018-0626-6}, \href
  {https://ui.adsabs.harvard.edu/abs/2019NatAs...3..167D} {3, 167}

\bibitem[\protect\citeauthoryear{{Donati} et~al.,}{{Donati}
  et~al.}{2011}]{Donati2011}
{Donati} J.~F.,  et~al., 2011, \mn@doi [\mnras]
  {10.1111/j.1365-2966.2011.19366.x}, \href
  {https://ui.adsabs.harvard.edu/abs/2011MNRAS.417.1747D} {417, 1747}

\bibitem[\protect\citeauthoryear{{Dubrulle}, {Morfill}  \&
  {Sterzik}}{{Dubrulle} et~al.}{1995}]{Dubrulle1995}
{Dubrulle} B.,  {Morfill} G.,   {Sterzik} M.,  1995, \mn@doi [\icarus]
  {10.1006/icar.1995.1058}, \href
  {https://ui.adsabs.harvard.edu/abs/1995Icar..114..237D} {114, 237}

\bibitem[\protect\citeauthoryear{{Dullemond}, {Juhasz}, {Pohl}, {Sereshti},
  {Shetty}, {Peters}, {Commercon}  \& {Flock}}{{Dullemond}
  et~al.}{2012}]{Dullemond2012}
{Dullemond} C.~P.,  {Juhasz} A.,  {Pohl} A.,  {Sereshti} F.,  {Shetty} R.,
  {Peters} T.,  {Commercon} B.,   {Flock} M.,  2012, {RADMC-3D: A multi-purpose
  radiative transfer tool} (\mn@eprint {ascl} {1202.015})

\bibitem[\protect\citeauthoryear{{Dullemond} et~al.,}{{Dullemond}
  et~al.}{2018}]{Dullemond2018}
{Dullemond} C.~P.,  et~al., 2018, \mn@doi [\apjl] {10.3847/2041-8213/aaf742},
  \href {https://ui.adsabs.harvard.edu/abs/2018ApJ...869L..46D} {869, L46}

\bibitem[\protect\citeauthoryear{{Dvorak}, {Froeschle}  \&
  {Froeschle}}{{Dvorak} et~al.}{1989}]{Dvorak1989}
{Dvorak} R.,  {Froeschle} C.,   {Froeschle} C.,  1989, \aap, \href
  {https://ui.adsabs.harvard.edu/abs/1989A&A...226..335D} {226, 335}

\bibitem[\protect\citeauthoryear{{Flaherty} et~al.,}{{Flaherty}
  et~al.}{2020}]{Flaherty2020}
{Flaherty} K.,  et~al., 2020, \mn@doi [\apj] {10.3847/1538-4357/ab8cc5}, \href
  {https://ui.adsabs.harvard.edu/abs/2020ApJ...895..109F} {895, 109}

\bibitem[\protect\citeauthoryear{{Flock}, {Ruge}, {Dzyurkevich}, {Henning},
  {Klahr}  \& {Wolf}}{{Flock} et~al.}{2015}]{Flock2015}
{Flock} M.,  {Ruge} J.~P.,  {Dzyurkevich} N.,  {Henning} T.,  {Klahr} H.,
  {Wolf} S.,  2015, \mn@doi [\aap] {10.1051/0004-6361/201424693}, \href
  {https://ui.adsabs.harvard.edu/abs/2015A&A...574A..68F} {574, A68}

\bibitem[\protect\citeauthoryear{{Fung}, {Zhu}  \& {Chiang}}{{Fung}
  et~al.}{2019}]{Fung2019}
{Fung} J.,  {Zhu} Z.,   {Chiang} E.,  2019, \mn@doi [\apj]
  {10.3847/1538-4357/ab53da}, \href
  {https://ui.adsabs.harvard.edu/abs/2019ApJ...887..152F} {887, 152}

\bibitem[\protect\citeauthoryear{{Gaia Collaboration}, {Brown}, {Vallenari},
  {Prusti}, {de Bruijne}, {Babusiaux}  \& {Biermann}}{{Gaia Collaboration}
  et~al.}{2020}]{Gaia2020}
{Gaia Collaboration} {Brown} A.~G.~A.,  {Vallenari} A.,  {Prusti} T.,  {de
  Bruijne} J.~H.~J.,  {Babusiaux} C.,   {Biermann} M.,  2020, arXiv e-prints,
  \href {https://ui.adsabs.harvard.edu/abs/2020arXiv201201533G} {p.
  arXiv:2012.01533}

\bibitem[\protect\citeauthoryear{{Haffert}, {Bohn}, {de Boer}, {Snellen},
  {Brinchmann}, {Girard}, {Keller}  \& {Bacon}}{{Haffert}
  et~al.}{2019}]{Haffert2019}
{Haffert} S.~Y.,  {Bohn} A.~J.,  {de Boer} J.,  {Snellen} I.~A.~G.,
  {Brinchmann} J.,  {Girard} J.~H.,  {Keller} C.~U.,   {Bacon} R.,  2019,
  \mn@doi [Nature Astronomy] {10.1038/s41550-019-0780-5}, \href
  {https://ui.adsabs.harvard.edu/abs/2019NatAs...3..749H} {3, 749}

\bibitem[\protect\citeauthoryear{{Haugb{\o}lle}, {Weber}, {Wielandt},
  {Ben{\'\i}tez-Llambay}, {Bizzarro}, {Gressel}  \& {Pessah}}{{Haugb{\o}lle}
  et~al.}{2019}]{Haugbolle2019}
{Haugb{\o}lle} T.,  {Weber} P.,  {Wielandt} D.~P.,  {Ben{\'\i}tez-Llambay} P.,
  {Bizzarro} M.,  {Gressel} O.,   {Pessah} M.~E.,  2019, \mn@doi [\aj]
  {10.3847/1538-3881/ab1591}, \href
  {https://ui.adsabs.harvard.edu/abs/2019AJ....158...55H} {158, 55}

\bibitem[\protect\citeauthoryear{{Holman} \& {Wiegert}}{{Holman} \&
  {Wiegert}}{1999}]{Holman1999}
{Holman} M.~J.,  {Wiegert} P.~A.,  1999, \mn@doi [\aj] {10.1086/300695}, \href
  {https://ui.adsabs.harvard.edu/abs/1999AJ....117..621H} {117, 621}

\bibitem[\protect\citeauthoryear{{Hu}, {Zhu}, {Okuzumi}, {Bai}, {Wang},
  {Tomida}  \& {Stone}}{{Hu} et~al.}{2019}]{Hu2019}
{Hu} X.,  {Zhu} Z.,  {Okuzumi} S.,  {Bai} X.-N.,  {Wang} L.,  {Tomida} K.,
  {Stone} J.~M.,  2019, \mn@doi [\apj] {10.3847/1538-4357/ab44cb}, \href
  {https://ui.adsabs.harvard.edu/abs/2019ApJ...885...36H} {885, 36}

\bibitem[\protect\citeauthoryear{Hunter}{Hunter}{2007}]{Matplotlib}
Hunter J.~D.,  2007, \mn@doi [Computing In Science \& Engineering]
  {10.1109/MCSE.2007.55}, 9, 90

\bibitem[\protect\citeauthoryear{{Izquierdo}, {Testi}, {Facchini}, {Rosotti}
  \& {van Dishoeck}}{{Izquierdo} et~al.}{2021}]{Izquierdo2021}
{Izquierdo} A.~F.,  {Testi} L.,  {Facchini} S.,  {Rosotti} G.~P.,   {van
  Dishoeck} E.~F.,  2021, \mn@doi [\aap] {10.1051/0004-6361/202140779}, \href
  {https://ui.adsabs.harvard.edu/abs/2021A&A...650A.179I} {650, A179}

\bibitem[\protect\citeauthoryear{{Johansen}, {Youdin}  \& {Klahr}}{{Johansen}
  et~al.}{2009}]{Johansen2009}
{Johansen} A.,  {Youdin} A.,   {Klahr} H.,  2009, \mn@doi [\apj]
  {10.1088/0004-637X/697/2/1269}, \href
  {https://ui.adsabs.harvard.edu/abs/2009ApJ...697.1269J} {697, 1269}

\bibitem[\protect\citeauthoryear{{Kanagawa}, {Muto}, {Tanaka}, {Tanigawa},
  {Takeuchi}, {Tsukagoshi}  \& {Momose}}{{Kanagawa}
  et~al.}{2016}]{Kanagawa2016}
{Kanagawa} K.~D.,  {Muto} T.,  {Tanaka} H.,  {Tanigawa} T.,  {Takeuchi} T.,
  {Tsukagoshi} T.,   {Momose} M.,  2016, \mn@doi [\pasj] {10.1093/pasj/psw037},
  \href {https://ui.adsabs.harvard.edu/abs/2016PASJ...68...43K} {68, 43}

\bibitem[\protect\citeauthoryear{{Keppler} et~al.,}{{Keppler}
  et~al.}{2018}]{Keppler2018}
{Keppler} M.,  et~al., 2018, \mn@doi [\aap] {10.1051/0004-6361/201832957},
  \href {https://ui.adsabs.harvard.edu/abs/2018A&A...617A..44K} {617, A44}

\bibitem[\protect\citeauthoryear{{Kley} \& {Haghighipour}}{{Kley} \&
  {Haghighipour}}{2014}]{Kley2014}
{Kley} W.,  {Haghighipour} N.,  2014, \mn@doi [\aap]
  {10.1051/0004-6361/201323235}, \href
  {https://ui.adsabs.harvard.edu/abs/2014A&A...564A..72K} {564, A72}

\bibitem[\protect\citeauthoryear{{Kurucz}}{{Kurucz}}{1979}]{Kurucz1979}
{Kurucz} R.~L.,  1979, \mn@doi [\apjs] {10.1086/190589}, \href
  {https://ui.adsabs.harvard.edu/abs/1979ApJS...40....1K} {40, 1}

\bibitem[\protect\citeauthoryear{{Lin}}{{Lin}}{2019}]{Lin2019}
{Lin} M.-K.,  2019, \mn@doi [\mnras] {10.1093/mnras/stz701}, \href
  {https://ui.adsabs.harvard.edu/abs/2019MNRAS.485.5221L} {485, 5221}

\bibitem[\protect\citeauthoryear{{Mamajek} \& {Bell}}{{Mamajek} \&
  {Bell}}{2014}]{Mamajek2014}
{Mamajek} E.~E.,  {Bell} C. P.~M.,  2014, \mn@doi [\mnras]
  {10.1093/mnras/stu1894}, \href
  {https://ui.adsabs.harvard.edu/abs/2014MNRAS.445.2169M} {445, 2169}

\bibitem[\protect\citeauthoryear{{Marino}, {Perez}  \& {Casassus}}{{Marino}
  et~al.}{2015}]{Marino2015}
{Marino} S.,  {Perez} S.,   {Casassus} S.,  2015, \mn@doi [\apjl]
  {10.1088/2041-8205/798/2/L44}, \href
  {https://ui.adsabs.harvard.edu/abs/2015ApJ...798L..44M} {798, L44}

\bibitem[\protect\citeauthoryear{{Martin}}{{Martin}}{2018}]{Martin2018}
{Martin} D.~V.,  2018, {Populations of Planets in Multiple Star Systems}.
p.~156, \mn@doi{10.1007/978-3-319-55333-7\_156}

\bibitem[\protect\citeauthoryear{{Martin}, {Zhu}  \& {Armitage}}{{Martin}
  et~al.}{2020}]{Martin2020}
{Martin} R.~G.,  {Zhu} Z.,   {Armitage} P.~J.,  2020, \mn@doi [\apjl]
  {10.3847/2041-8213/aba3c1}, \href
  {https://ui.adsabs.harvard.edu/abs/2020ApJ...898L..26M} {898, L26}

\bibitem[\protect\citeauthoryear{{Martin}, {Zhu}, {Armitage}, {Yang}  \&
  {Baehr}}{{Martin} et~al.}{2021}]{Martin2021}
{Martin} R.~G.,  {Zhu} Z.,  {Armitage} P.~J.,  {Yang} C.-C.,   {Baehr} H.,
  2021, \mn@doi [\mnras] {10.1093/mnras/stab232}, \href
  {https://ui.adsabs.harvard.edu/abs/2021MNRAS.502.4426M} {502, 4426}

\bibitem[\protect\citeauthoryear{{Martinez-Brunner} et~al.,}{{Martinez-Brunner}
  et~al.}{2021}]{Martinez2021}
{Martinez-Brunner} R.,  et~al., 2021, subm.

\bibitem[\protect\citeauthoryear{{Masset}}{{Masset}}{2000}]{Masset2000}
{Masset} F.,  2000, \mn@doi [\aaps] {10.1051/aas:2000116}, \href
  {https://ui.adsabs.harvard.edu/abs/2000A&AS..141..165M} {141, 165}

\bibitem[\protect\citeauthoryear{{McMullin}, {Waters}, {Schiebel}, {Young}  \&
  {Golap}}{{McMullin} et~al.}{2007}]{McMullin2007}
{McMullin} J.~P.,  {Waters} B.,  {Schiebel} D.,  {Young} W.,   {Golap} K.,
  2007, in {Shaw} R.~A.,  {Hill} F.,   {Bell} D.~J.,  eds,  Astronomical
  Society of the Pacific Conference Series Vol. 376, Astronomical Data Analysis
  Software and Systems XVI. p.~127

\bibitem[\protect\citeauthoryear{{Min}, {Rab}, {Woitke}, {Dominik}  \&
  {M{\'e}nard}}{{Min} et~al.}{2016}]{Min2016}
{Min} M.,  {Rab} C.,  {Woitke} P.,  {Dominik} C.,   {M{\'e}nard} F.,  2016,
  \mn@doi [\aap] {10.1051/0004-6361/201526048}, \href
  {https://ui.adsabs.harvard.edu/abs/2016A&A...585A..13M} {585, A13}

\bibitem[\protect\citeauthoryear{{Montesinos}, {Garrido-Deutelmoser},
  {Olofsson}, {Giuppone}, {Cuadra}, {Bayo}, {Sucerquia}  \&
  {Cuello}}{{Montesinos} et~al.}{2020}]{Montesinos2020}
{Montesinos} M.,  {Garrido-Deutelmoser} J.,  {Olofsson} J.,  {Giuppone} C.~A.,
  {Cuadra} J.,  {Bayo} A.,  {Sucerquia} M.,   {Cuello} N.,  2020, \mn@doi
  [\aap] {10.1051/0004-6361/202038758}, \href
  {https://ui.adsabs.harvard.edu/abs/2020A&A...642A.224M} {642, A224}

\bibitem[\protect\citeauthoryear{{Montesinos}, {Cuello}, {Olofsson}, {Cuadra},
  {Bayo}, {Bertrang}  \& {Perrot}}{{Montesinos} et~al.}{2021}]{Montesinos2021}
{Montesinos} M.,  {Cuello} N.,  {Olofsson} J.,  {Cuadra} J.,  {Bayo} A.,
  {Bertrang} G. H.~M.,   {Perrot} C.,  2021, arXiv e-prints, \href
  {https://ui.adsabs.harvard.edu/abs/2021arXiv210202874M} {p. arXiv:2102.02874}

\bibitem[\protect\citeauthoryear{{M{\"u}ller} et~al.,}{{M{\"u}ller}
  et~al.}{2018}]{Muller2018}
{M{\"u}ller} A.,  et~al., 2018, \mn@doi [\aap] {10.1051/0004-6361/201833584},
  \href {https://ui.adsabs.harvard.edu/abs/2018A&A...617L...2M} {617, L2}

\bibitem[\protect\citeauthoryear{{Penzlin}, {Kley}  \& {Nelson}}{{Penzlin}
  et~al.}{2021}]{Penzlin2021}
{Penzlin} A. B.~T.,  {Kley} W.,   {Nelson} R.~P.,  2021, \mn@doi [\aap]
  {10.1051/0004-6361/202039319}, \href
  {https://ui.adsabs.harvard.edu/abs/2021A&A...645A..68P} {645, A68}

\bibitem[\protect\citeauthoryear{Perez \& Granger}{Perez \&
  Granger}{2007}]{ipython}
Perez F.,  Granger B.~E.,  2007, \mn@doi [Computing in Science and Engg.]
  {10.1109/MCSE.2007.53}, 9, 21

\bibitem[\protect\citeauthoryear{{Perez}, {Dunhill}, {Casassus}, {Roman},
  {Szul{\'a}gyi}, {Flores}, {Marino}  \& {Montesinos}}{{Perez}
  et~al.}{2015}]{Perez2015}
{Perez} S.,  {Dunhill} A.,  {Casassus} S.,  {Roman} P.,  {Szul{\'a}gyi} J.,
  {Flores} C.,  {Marino} S.,   {Montesinos} M.,  2015, \mn@doi [\apjl]
  {10.1088/2041-8205/811/1/L5}, \href
  {https://ui.adsabs.harvard.edu/abs/2015ApJ...811L...5P} {811, L5}

\bibitem[\protect\citeauthoryear{{P{\'e}rez}, {Casassus}  \&
  {Ben{\'\i}tez-Llambay}}{{P{\'e}rez} et~al.}{2018}]{Perez2018}
{P{\'e}rez} S.,  {Casassus} S.,   {Ben{\'\i}tez-Llambay} P.,  2018, \mn@doi
  [\mnras] {10.1093/mnrasl/sly109}, \href
  {https://ui.adsabs.harvard.edu/abs/2018MNRAS.480L..12P} {480, L12}

\bibitem[\protect\citeauthoryear{{Pierens} \& {Nelson}}{{Pierens} \&
  {Nelson}}{2013}]{Pierens2013}
{Pierens} A.,  {Nelson} R.~P.,  2013, \mn@doi [\aap]
  {10.1051/0004-6361/201321777}, \href
  {https://ui.adsabs.harvard.edu/abs/2013A&A...556A.134P} {556, A134}

\bibitem[\protect\citeauthoryear{{Pinte} et~al.,}{{Pinte}
  et~al.}{2018}]{Pinte2018}
{Pinte} C.,  et~al., 2018, \mn@doi [\apjl] {10.3847/2041-8213/aac6dc}, \href
  {https://ui.adsabs.harvard.edu/abs/2018ApJ...860L..13P} {860, L13}

\bibitem[\protect\citeauthoryear{{Pinte} et~al.,}{{Pinte}
  et~al.}{2020}]{Pinte2020}
{Pinte} C.,  et~al., 2020, \mn@doi [\apjl] {10.3847/2041-8213/ab6dda}, \href
  {https://ui.adsabs.harvard.edu/abs/2020ApJ...890L...9P} {890, L9}

\bibitem[\protect\citeauthoryear{{Quast}, {Torres}, {de La Reza}, {da Silva}
  \& {Mayor}}{{Quast} et~al.}{2000}]{Quast2000}
{Quast} G.~R.,  {Torres} C. A.~O.,  {de La Reza} R.,  {da Silva} L.,   {Mayor}
  M.,  2000, IAU Symposium, \href
  {https://ui.adsabs.harvard.edu/abs/2000IAUS..200P..28Q} {200, 28}

\bibitem[\protect\citeauthoryear{{Rapson}, {Kastner}, {Andrews}, {Hines},
  {Macintosh}, {Millar-Blanchaer}  \& {Tamura}}{{Rapson}
  et~al.}{2015}]{Rapson2015}
{Rapson} V.~A.,  {Kastner} J.~H.,  {Andrews} S.~M.,  {Hines} D.~C.,
  {Macintosh} B.,  {Millar-Blanchaer} M.,   {Tamura} M.,  2015, \mn@doi [\apjl]
  {10.1088/2041-8205/803/1/L10}, \href
  {https://ui.adsabs.harvard.edu/abs/2015ApJ...803L..10R} {803, L10}

\bibitem[\protect\citeauthoryear{{Rodenkirch}, {Rometsch}, {Dullemond}, {Weber}
   \& {Kley}}{{Rodenkirch} et~al.}{2021}]{Rodenkirch2021}
{Rodenkirch} P.~J.,  {Rometsch} T.,  {Dullemond} C.~P.,  {Weber} P.,   {Kley}
  W.,  2021, \mn@doi [\aap] {10.1051/0004-6361/202038484}, \href
  {https://ui.adsabs.harvard.edu/abs/2021A&A...647A.174R} {647, A174}

\bibitem[\protect\citeauthoryear{{Rosenfeld}, {Andrews}, {Wilner}  \&
  {Stempels}}{{Rosenfeld} et~al.}{2012}]{Rosenfeld2012}
{Rosenfeld} K.~A.,  {Andrews} S.~M.,  {Wilner} D.~J.,   {Stempels} H.~C.,
  2012, \mn@doi [\apj] {10.1088/0004-637X/759/2/119}, \href
  {https://ui.adsabs.harvard.edu/abs/2012ApJ...759..119R} {759, 119}

\bibitem[\protect\citeauthoryear{{Rosenfeld}, {Andrews}, {Wilner}, {Kastner}
  \& {McClure}}{{Rosenfeld} et~al.}{2013}]{Rosenfeld2013}
{Rosenfeld} K.~A.,  {Andrews} S.~M.,  {Wilner} D.~J.,  {Kastner} J.~H.,
  {McClure} M.~K.,  2013, \mn@doi [\apj] {10.1088/0004-637X/775/2/136}, \href
  {https://ui.adsabs.harvard.edu/abs/2013ApJ...775..136R} {775, 136}

\bibitem[\protect\citeauthoryear{{Rosotti}, {Juhasz}, {Booth}  \&
  {Clarke}}{{Rosotti} et~al.}{2016}]{Rosotti2016}
{Rosotti} G.~P.,  {Juhasz} A.,  {Booth} R.~A.,   {Clarke} C.~J.,  2016, \mn@doi
  [\mnras] {10.1093/mnras/stw691}, \href
  {https://ui.adsabs.harvard.edu/abs/2016MNRAS.459.2790R} {459, 2790}

\bibitem[\protect\citeauthoryear{{Ru{\'\i}z-Rodr{\'\i}guez}, {Kastner}, {Dong},
  {Principe}, {Andrews}  \& {Wilner}}{{Ru{\'\i}z-Rodr{\'\i}guez}
  et~al.}{2019}]{Ruiz-Rodriguez2019}
{Ru{\'\i}z-Rodr{\'\i}guez} D.,  {Kastner} J.~H.,  {Dong} R.,  {Principe} D.~A.,
   {Andrews} S.~M.,   {Wilner} D.~J.,  2019, \mn@doi [\aj]
  {10.3847/1538-3881/ab1c58}, \href
  {https://ui.adsabs.harvard.edu/abs/2019AJ....157..237R} {157, 237}

\bibitem[\protect\citeauthoryear{{Saito} \& {Sirono}}{{Saito} \&
  {Sirono}}{2011}]{Saito2011}
{Saito} E.,  {Sirono} S.-i.,  2011, \mn@doi [\apj]
  {10.1088/0004-637X/728/1/20}, \href
  {https://ui.adsabs.harvard.edu/abs/2011ApJ...728...20S} {728, 20}

\bibitem[\protect\citeauthoryear{{Schulik}, {Johansen}, {Bitsch}  \&
  {Lega}}{{Schulik} et~al.}{2019}]{Schulik2019}
{Schulik} M.,  {Johansen} A.,  {Bitsch} B.,   {Lega} E.,  2019, \mn@doi [\aap]
  {10.1051/0004-6361/201935473}, \href
  {https://ui.adsabs.harvard.edu/abs/2019A&A...632A.118S} {632, A118}

\bibitem[\protect\citeauthoryear{{Shakura} \& {Sunyaev}}{{Shakura} \&
  {Sunyaev}}{1973}]{Shakura1973}
{Shakura} N.~I.,  {Sunyaev} R.~A.,  1973, \aap, \href
  {https://ui.adsabs.harvard.edu/abs/1973A&A....24..337S} {500, 33}

\bibitem[\protect\citeauthoryear{{Simon} \& {Armitage}}{{Simon} \&
  {Armitage}}{2014}]{Simon2014}
{Simon} J.~B.,  {Armitage} P.~J.,  2014, \mn@doi [\apj]
  {10.1088/0004-637X/784/1/15}, \href
  {https://ui.adsabs.harvard.edu/abs/2014ApJ...784...15S} {784, 15}

\bibitem[\protect\citeauthoryear{{Stolker}, {Dominik}, {Min}, {Garufi},
  {Mulders}  \& {Avenhaus}}{{Stolker} et~al.}{2016}]{Stolker2016}
{Stolker} T.,  {Dominik} C.,  {Min} M.,  {Garufi} A.,  {Mulders} G.~D.,
  {Avenhaus} H.,  2016, \mn@doi [\aap] {10.1051/0004-6361/201629098}, \href
  {https://ui.adsabs.harvard.edu/abs/2016A&A...596A..70S} {596, A70}

\bibitem[\protect\citeauthoryear{{Sutherland} \& {Kratter}}{{Sutherland} \&
  {Kratter}}{2019}]{Sutherland2019}
{Sutherland} A.~P.,  {Kratter} K.~M.,  2019, \mn@doi [\mnras]
  {10.1093/mnras/stz1503}, \href
  {https://ui.adsabs.harvard.edu/abs/2019MNRAS.487.3288S} {487, 3288}

\bibitem[\protect\citeauthoryear{{Szul{\'a}gyi}}{{Szul{\'a}gyi}}{2017}]{Szulagyi2017}
{Szul{\'a}gyi} J.,  2017, \mn@doi [\apj] {10.3847/1538-4357/aa7515}, \href
  {https://ui.adsabs.harvard.edu/abs/2017ApJ...842..103S} {842, 103}

\bibitem[\protect\citeauthoryear{{Szul{\'a}gyi}, {Mayer}  \&
  {Quinn}}{{Szul{\'a}gyi} et~al.}{2017}]{Szulagyi2017b}
{Szul{\'a}gyi} J.,  {Mayer} L.,   {Quinn} T.,  2017, \mn@doi [\mnras]
  {10.1093/mnras/stw2617}, \href
  {https://ui.adsabs.harvard.edu/abs/2017MNRAS.464.3158S} {464, 3158}

\bibitem[\protect\citeauthoryear{{Szul{\'a}gyi}, {Plas}, {Meyer}, {Pohl},
  {Quanz}, {Mayer}, {Daemgen}  \& {Tamburello}}{{Szul{\'a}gyi}
  et~al.}{2018}]{Szulagyi2018}
{Szul{\'a}gyi} J.,  {Plas} G. v.~d.,  {Meyer} M.~R.,  {Pohl} A.,  {Quanz}
  S.~P.,  {Mayer} L.,  {Daemgen} S.,   {Tamburello} V.,  2018, \mn@doi [\mnras]
  {10.1093/mnras/stx2602}, \href
  {https://ui.adsabs.harvard.edu/abs/2018MNRAS.473.3573S} {473, 3573}

\bibitem[\protect\citeauthoryear{{Szul{\'a}gyi}, {Binkert}  \&
  {Surville}}{{Szul{\'a}gyi} et~al.}{2021}]{Szulagyi2021}
{Szul{\'a}gyi} J.,  {Binkert} F.,   {Surville} C.,  2021, arXiv e-prints, \href
  {https://ui.adsabs.harvard.edu/abs/2021arXiv210312128S} {p. arXiv:2103.12128}

\bibitem[\protect\citeauthoryear{{Teague}, {Bae}, {Bergin}, {Birnstiel}  \&
  {Foreman-Mackey}}{{Teague} et~al.}{2018}]{Teague2018}
{Teague} R.,  {Bae} J.,  {Bergin} E.~A.,  {Birnstiel} T.,   {Foreman-Mackey}
  D.,  2018, \mn@doi [\apjl] {10.3847/2041-8213/aac6d7}, \href
  {https://ui.adsabs.harvard.edu/abs/2018ApJ...860L..12T} {860, L12}

\bibitem[\protect\citeauthoryear{{Teague}, {Bae}  \& {Bergin}}{{Teague}
  et~al.}{2019}]{Teague2019}
{Teague} R.,  {Bae} J.,   {Bergin} E.~A.,  2019, \mn@doi [\nat]
  {10.1038/s41586-019-1642-0}, \href
  {https://ui.adsabs.harvard.edu/abs/2019Natur.574..378T} {574, 378}

\bibitem[\protect\citeauthoryear{{Weber}, {Ben{\'\i}tez-Llambay}, {Gressel},
  {Krapp}  \& {Pessah}}{{Weber} et~al.}{2018}]{Weber2018}
{Weber} P.,  {Ben{\'\i}tez-Llambay} P.,  {Gressel} O.,  {Krapp} L.,   {Pessah}
  M.~E.,  2018, \mn@doi [\apj] {10.3847/1538-4357/aaab63}, \href
  {https://ui.adsabs.harvard.edu/abs/2018ApJ...854..153W} {854, 153}

\bibitem[\protect\citeauthoryear{{Weber}, {P{\'e}rez}, {Ben{\'\i}tez-Llambay},
  {Gressel}, {Casassus}  \& {Krapp}}{{Weber} et~al.}{2019}]{Weber2019}
{Weber} P.,  {P{\'e}rez} S.,  {Ben{\'\i}tez-Llambay} P.,  {Gressel} O.,
  {Casassus} S.,   {Krapp} L.,  2019, \mn@doi [\apj]
  {10.3847/1538-4357/ab412f}, \href
  {https://ui.adsabs.harvard.edu/abs/2019ApJ...884..178W} {884, 178}

\bibitem[\protect\citeauthoryear{{Zhang}, {Blake}  \& {Bergin}}{{Zhang}
  et~al.}{2015}]{Zhang2015}
{Zhang} K.,  {Blake} G.~A.,   {Bergin} E.~A.,  2015, \mn@doi [\apjl]
  {10.1088/2041-8205/806/1/L7}, \href
  {https://ui.adsabs.harvard.edu/abs/2015ApJ...806L...7Z} {806, L7}

\bibitem[\protect\citeauthoryear{{Zhang} et~al.,}{{Zhang}
  et~al.}{2018}]{Zhang2018}
{Zhang} S.,  et~al., 2018, \mn@doi [\apjl] {10.3847/2041-8213/aaf744}, \href
  {https://ui.adsabs.harvard.edu/abs/2018ApJ...869L..47Z} {869, L47}

\bibitem[\protect\citeauthoryear{{Zhu}}{{Zhu}}{2015}]{Zhu2015a}
{Zhu} Z.,  2015, \mn@doi [\apj] {10.1088/0004-637X/799/1/16}, \href
  {https://ui.adsabs.harvard.edu/abs/2015ApJ...799...16Z} {799, 16}

\bibitem[\protect\citeauthoryear{{Zhu}, {Nelson}, {Hartmann}, {Espaillat}  \&
  {Calvet}}{{Zhu} et~al.}{2011}]{Zhu2011}
{Zhu} Z.,  {Nelson} R.~P.,  {Hartmann} L.,  {Espaillat} C.,   {Calvet} N.,
  2011, \mn@doi [\apj] {10.1088/0004-637X/729/1/47}, \href
  {https://ui.adsabs.harvard.edu/abs/2011ApJ...729...47Z} {729, 47}

\bibitem[\protect\citeauthoryear{{Zhu}, {Stone}  \& {Bai}}{{Zhu}
  et~al.}{2015}]{Zhu2015b}
{Zhu} Z.,  {Stone} J.~M.,   {Bai} X.-N.,  2015, \mn@doi [\apj]
  {10.1088/0004-637X/801/2/81}, \href
  {https://ui.adsabs.harvard.edu/abs/2015ApJ...801...81Z} {801, 81}

\bibitem[\protect\citeauthoryear{{van Holstein} et~al.,}{{van Holstein}
  et~al.}{2020}]{vanHolstein2020}
{van Holstein} R.~G.,  et~al., 2020, \mn@doi [\aap]
  {10.1051/0004-6361/201834996}, \href
  {https://ui.adsabs.harvard.edu/abs/2020A&A...633A..64V} {633, A64}

\bibitem[\protect\citeauthoryear{van~der Walt, Chris~Colbert  \&
  Varoquaux}{van~der Walt et~al.}{2011}]{numpy}
van~der Walt S.,  Chris~Colbert S.,   Varoquaux G.,  2011, \mn@doi [Computing
  in Science \& Engineering] {10.1109/MCSE.2011.37}, 13, 22

\makeatother
\end{thebibliography}

\appendix

\section{Spectral Energy Distribution}\label{appendix:SED}
\begin{figure}
    \centering
    \includegraphics{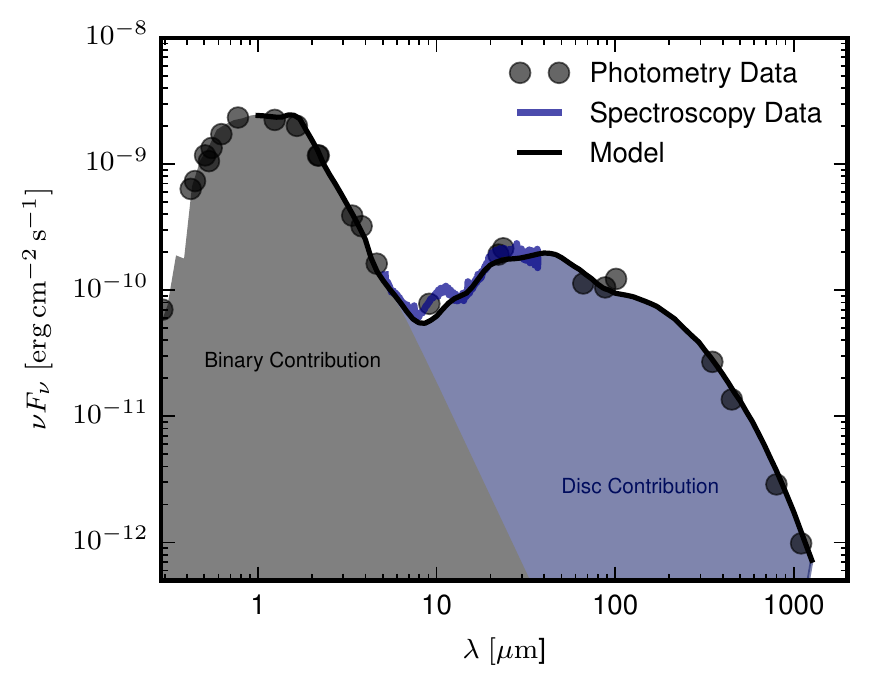}
    \caption{SED of the hydrodynamical+RT model. The contributions by direct stellar light and by the disc are highlighted.}
    \label{fig:SED}
\end{figure}

To adapt our disc model to the case of V4046~Sgr, the SED is an important input. A complete model has to recover the observed fluxes at all wavelengths to be consistent. In Fig.~\ref{fig:SED} we show the comparison of our model, resulting from hydrodynamics and radiative transfer, to the observed data from spectroscopy and photometry at recorded wavelengths. The contributions of  the binary stars and the disc material are highlighted. The references for the observational data are summarised in \citetalias[][]{Martinez2021}.

\section{Shadow of an inclined CPD}\label{appendix:inclined}
A perceivable shadow of the CPD is cast, if there is sufficient material high enough the PPD's mid-plane. This can be due to a very vertically inflated CPD as shown in Sec.~\ref{subsec:simulations}. Another possibility is to have a CPD that is inclined relative to the mid-plane of the surrounding PPD, which is a possible geometry for detached CPDs \citep[][]{Martin2020,Martin2021}. To test this, we reduce the vertical scale to $\Tilde{h} = 0.1$ and off-set the mid-plane of the CPD from the mid-plane of the PPD by an angle $i_{\rm CPD} = 60^{\circ}$. The axis of rotation is chosen to be perpendicular to the line that connects centre of CPD and centre of PPD. The maximum elevation of the CPD mid-plane is $H_{\max} = \sin{\left(i_{\rm CPD}\right)}R_{\rm CPD}$ or in terms of the aspect ratio, $h_{\max} = \sin{\left(i_{\rm CPD}\right)}R_{\rm CPD} r_{\rm p}^{-1}$. From our simulations we find that this relatively flat, inclined CPD around a {$3.7\,m_{\rm J}$} does not cast a visible shadow onto the bright ring. The vertical extend is not sufficient to affect the heights where scattered light originates. The shadow can only be recovered when increasing the companion's mass to about $m_{\rm P} = 16\,m_{\rm J}$ -- and hence increase the size of the parametric CPD by a factor of 1.63.

\section{Influence of two extended light sources for the shape of a shadow}\label{appendix:binary}
The calculation presented in Sec.~\ref{subsec:analysis} implicitly assumes one central, point-like source of radiation, as it makes use of the idea that the solid angle that the shadow covers is independent of its distance (both to the light source and to the ring), which implies that it corresponds to the solid angle covered by the CPD. This assumption breaks if the extent of the light source is of similar or larger size than the extent of the object. In V4046~Sgr the central binary is of close-to-equal constituents, with a separation of 0.041$\AU$. This is not trivially negligible. 
\begin{figure}
     \centering
     \begin{minipage}{\columnwidth}
         \centering
         \includegraphics[width=\columnwidth]{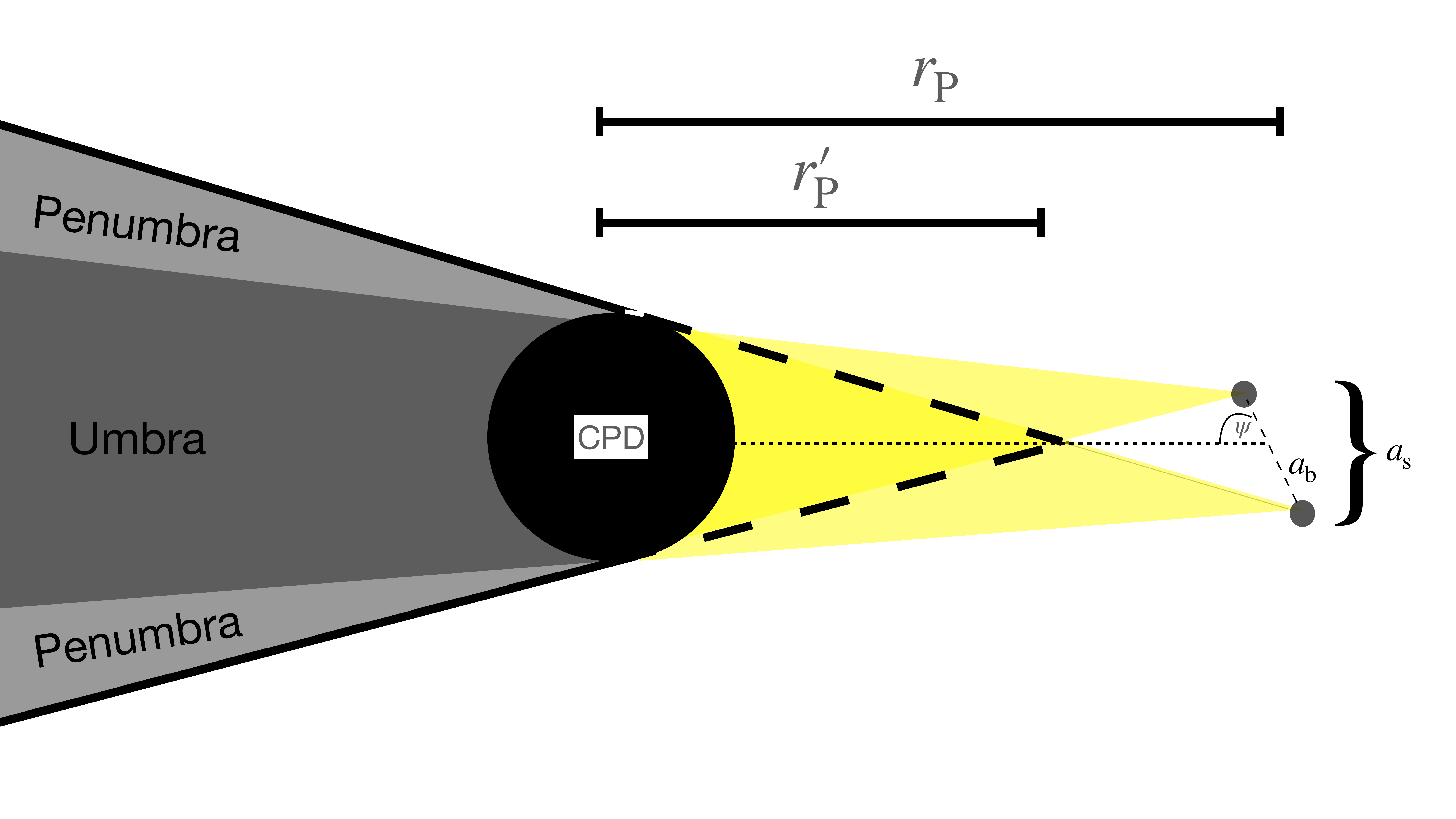}
     \end{minipage}
     \begin{minipage}{\columnwidth}
         \centering
         \includegraphics[width=\columnwidth]{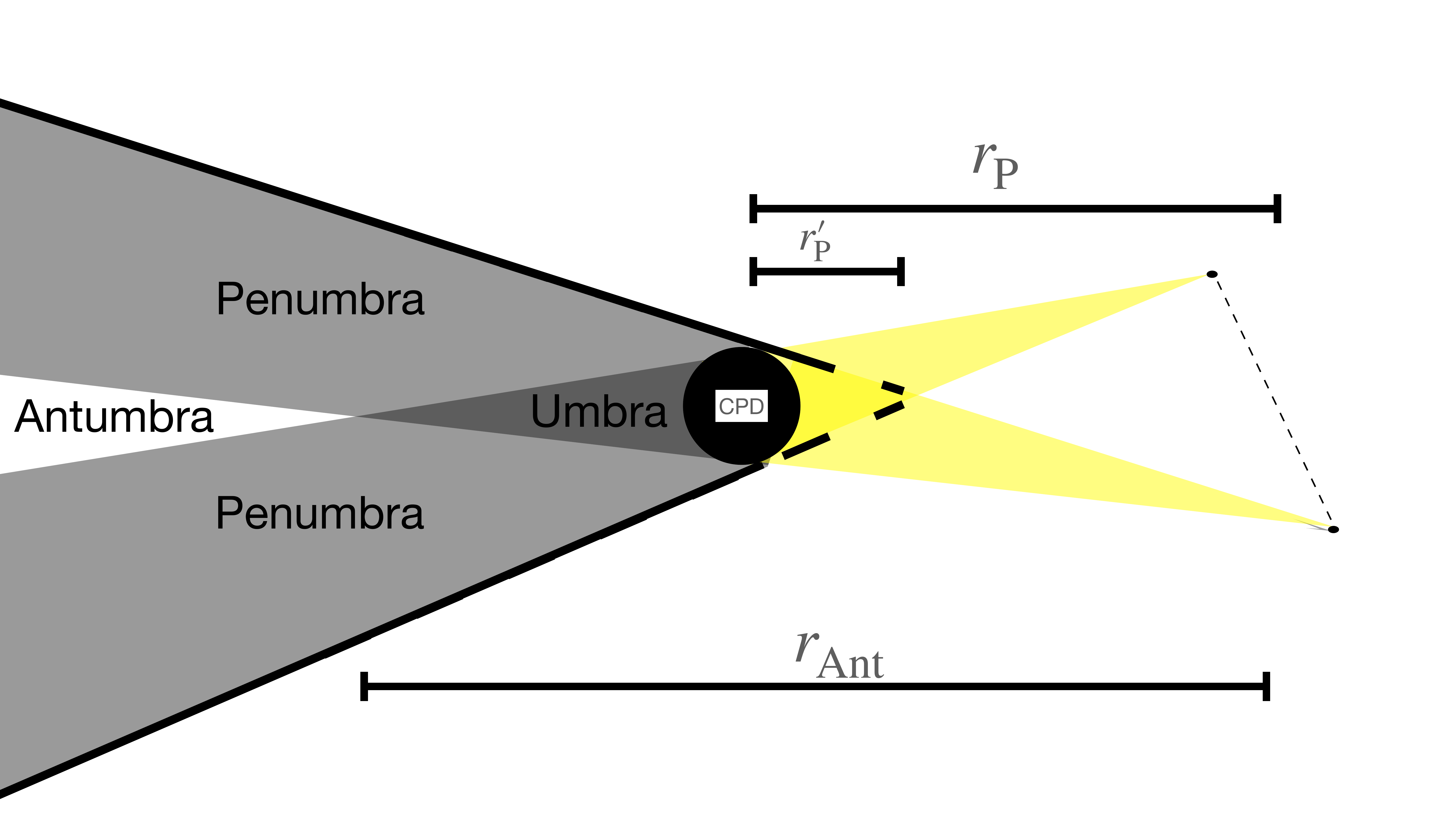}
     \end{minipage}
       \caption{CPD shadow in two different cases of illumination by binary. In the \textit{top panel} the source width, $a_{\rm s}$, is smaller than the extent of the CPD, creating a deep shadow (Umbra) flanked by two shadows that only correspond to one binary star (Penumbra). In the \textit{bottom panel} $a_{\rm s}$ is larger than the object, creating the Umbra only in a specific radial and azimuthal zone behind the CPD. Antumbra labels the region which is radially directly behind the CPD but exposed to light from both binary components due to the source separation.}
            \label{fig:binarysketch}
\end{figure}

We show a sketch of two specific scenarios in Fig.~\ref{fig:binarysketch}. The relevant width of the light source, $a_{\rm s}$, is the binary's separation projected into the plane of the shadow image, plus the contribution of the extent of the individual stellar components. Let $\psi$ be the angle between the planet's azimuthal location and the binary's orientation of conjunction (this means that $\psi=0$ when all three bodies are in one line). 
The maximum separation of radiation relevant for the shadowing is given by the \textit{source width}
$a_{\rm s} =  a_{\rm b} |\sin \psi|$. 
Looking at Fig.~\ref{fig:binarysketch}, we see that there is a focal point where the light cones of both stars meet. The focal point is at a distance to the planet of
\begin{equation}
    r'_{\rm P} = r_{\rm P}\frac{R_{\rm CPD}}{R_{\rm CPD} +a_{\rm s}/2}\,.
\end{equation}
The binary light source describes a magnified opening angle for the total extent of the CPD shadow, compared with a single light source:
\begin{equation}\label{equ:delta-prime}
    \Delta \varphi'_{\rm CPD} \approx \frac{d_{\rm CPD}}{r'_{\rm P}} = \frac{d_{\rm CPD} + a_{\rm s}}{r_{\rm P}}\,.
\end{equation} 
As mentioned in the previous section, in the case of a non-extended light source this opening angle is simply $\Delta \varphi_{\rm CPD} \approx d_{\rm CPD}/r_{\rm P}$. Therefore, one can identify an angular magnification factor of the signal:
\begin{equation}\label{equ:magnification}
    M = \frac{\Delta \varphi'_{\rm CPD}}{\Delta \varphi_{\rm CPD}} = 1+ \frac{a_{\rm s}}{d_{\rm CPD}}\,.
\end{equation}
As the size of the CPD is expected to increase with the planet's orbital distance, we see that this magnification factor becomes more and more irrelevant for objects on large orbits, but may well be relevant for objects close to the centre. 

We emphasise, that even though the shadow becomes larger under this magnification, the integrated decrement's equivalent width does not change. It is, therefore, not relevant for the details of the observed profile, as long as the shape of the magnified signal is dominated by the smoothing effect of the projected PSF. This can be used to formulate a maximally allowed magnification factor.

Further, if the width of the light source is larger than the width of the shadowing object, we can calculate a radial distance at which the Antumbra (i.e. the locations from which the angular width of the obscuring object is entirely contained within the angular width of light source as shown in the bottom panel of Fig.~\ref{fig:binarysketch}) starts to appear. 
Let $r_{\rm Ant}$ be the distance to the centre of mass at which Umbra turns into Antumbra. From the intercept theorem we get the connection $a_{\rm s}/r_{\rm Ant} = d_{\rm CPD}/(r_{\rm Ant}-r_{\rm P})$ and thus,
\begin{equation}\label{equ:rant}
    r_{\rm Ant} = \frac{r_{\rm P}}{1-d_{\rm CPD}/a_{\rm s}}\,.
\end{equation}
Applying the intersect theorem again we find that for $r>r_{\rm Ant}$, the size of the Antumbra grows according to 
\begin{equation}
    d_{\rm Ant}(r)=\frac{a_{\rm s}}{r_{\rm Ant}}(r-r_{\rm Ant})\,,
\end{equation}
which means that the azimuthal angle covered by the Antumbra is given by 
\begin{equation}
    \Delta \varphi_{\rm Ant}(r) = 2{\rm arctan}\left(\frac{d_{\rm Ant}}{2r}\right)\,.
\end{equation} 
This feature can put further constraints on the extent of a shadowing object for cases in which the source separation, $a_{\rm s}$, is larger than the extent of the object itself. 
In V4046$\,$Sgr, the azimuthal location of the intensity decrement on the PPD's faint side is unfavourable to analyse this feature. It has to be seen in future observations whether an anomaly between the two rings in the scattered light image can be inferred.

Thanks to the work by \citet{DOrazi2019} we know the binary's orientation at the same moment that the decrement was observed to a high level of confidence. They provide the binary's phase for the VLT/SPHERE observation in March 2016 as $\varphi_{\rm bin}=171.5^\circ$.

Note, that due to the finite travelling time of light, the observed shadows reflect a past state of the inner binary system. The signal is delayed by roughly two hours ($12.4^\circ$) with respect to the inner system configuration that caused the shadows. 
For the analysis of the CPD shadow, this is not important as this temporal shift applies to all three observed shadows. Yet, the light reaching us from the far side of the disc takes slightly longer than the light coming from the near side. This creates a disparity between the inner system's configurations to which each shadow corresponds -- i.e. a shadow observed in the near side stems from a slightly older system (as its light took less time to reach us) than a shadow observed from the far side. The temporal difference with respect to the light coming from the central binary is given by 
\begin{equation}
\Delta t_{\rm signal} = \frac{r_{\rm ring}}{c}-\sin(i)\sin(\varphi)\frac{r_{\rm ring}}{c}\,.
\end{equation}
This equation is neglecting the vertical extent of the ring (for the full expression see \citealp{DOrazi2019}). Here, the first term corresponds to the light traveling time from the star to the ring, the second term accounts for the temporal difference due to the disc's inclination (as already pointed out by \citealp{DOrazi2019}). Effectively, this means that the intensity decrement observed on the far-side of the disc is older by about three hours than the shadow linked to the star labelled binary 1 in the left panel of Fig.~\ref{fig:az-cut} and older by half an hour for the star labelled binary 2. This is not negligible as the period of the binary is only 2.42 $\,$d. After correcting for this effect we infer a value of $\psi_1 = 0.80$ from comparing the azimuth of the putative CPD shadow to binary star 1 and a value of $\psi_2 = 0.95$ from binary star 2. We use the mean value of $\psi=0.875$ for this analysis. With this, we obtain a source separation of $a_{\rm s} = 0.0315\AU$ at the moment of shadowing.\\
To explicitly demonstrate the relevance of this, we take the example of a planet of $10\,{\rm M}_{\rm J}$ at a distance of $1\AU$, hosting a shadowing CPD of $d_{\rm CPD} = 2/3\,R_{\rm Hill} = 0.081\AU$. The magnification factor for this case would be $M=1.51$, so significantly increases the width of the CPD shadow. For the same planet but at a distance of $r_{\rm P} = 0.5\AU$ the CPD diameter would be equal to the source separation, $d_{\rm CPD}= 0.041\AU$, and the magnification factor, therefore, a value of two. At smaller radii an antumbra is created and, at a distance given in equation~(\ref{equ:rant}), the two individual shadows of the stars cease to fully overlap. As long as the total extent of the two shadows on the broadcasting ring is efficiently smoothed by the PSF of the telescope, this effect is purely academical. 
However, this allows us to estimate a minimum allowed distance, $r_{\rm P,min}$, such that the produced shadow stays within the limits of the PSF. The critical scenario is given where
\begin{equation}
    \Delta\varphi_{\rm PSF} \overset{!}{=} \Delta \varphi_{\rm CPD} = \frac{d_{\rm CPD}}{r_{\rm P,min}}M\,. 
\end{equation}
From this, we obtain the connection 
\begin{equation}
    R_{\rm CPD} = \Delta \varphi_{\rm PSF}r_{\rm P,min} -a_{\rm s}\,,
\end{equation}
and again assuming the relation $R_ {\rm CPD}= 0.4R_{\rm Hill}$:
\begin{equation}\label{equ:amin-visual}
    r_{\rm P,min} = a_{\rm s}\left[\Delta\varphi_{\rm PSF}-\frac{2}{5}\left(\frac{m_{\rm P}}{3m_{\rm b}}\right)^{1/3}\right]^{-1}\,.
\end{equation}

\bsp
\label{lastpage}
\end{document}